\documentclass[twocolumn,english]{revtex4-2}
\usepackage[T1]{fontenc}
\usepackage[latin9]{inputenc}
\usepackage{amsmath}
\usepackage{amssymb}
\usepackage{graphicx}
\usepackage{esint}
\usepackage{babel}
\usepackage{xcolor}

\begin{document}
\title{Su-Schrieffer-Heeger quasicrystal: Topology, localization, and mobility edge}
\author{D. A. Miranda$^{1}$,  T.  V.  C.   Ant{\~a}o$^{2}$, and N.M.R Peres$^{1,3,4}$}
\affiliation{$^{1}$Centro de Física (CF-UM-UP) and Departamento de Física, Universidade
do Minho, P-4710-057 Braga, Portugal }
\affiliation{$^{2}$Department of Applied Physics, Aalto University, 02150 Espoo, Finland}
\affiliation{ $^{3}$International Iberian
Nanotechnology Laboratory (INL), Av Mestre José Veiga, 4715-330 Braga,
Portugal }
\affiliation{ $^{4}$POLIMA-Center for Polariton-driven Light-Matter
Interactions, University of Southern Denmark, Campusvej 55, DK-5230
Odense M, Denmark}

\begin{abstract}
In this paper we discussed the topological transition between trivial
and nontrivial phases of a quasi-periodic (Aubry-André like) mechanical
Su-Schrieffer-Heeger (SSH) model. We find that there exists a nontrivial
boundary separating the two topological phases and an analytical expression
for this boundary is found. 
We discuss the localization of the vibrational modes using the calculation of the inverse participation ratio (IPR)
and access the localization nature of the states of the system. We find three different regimes: extended, localized, and critical, depending on the intensity of the Aubry-André spring.
We further study the energy dependent
mobility edge (ME) separating localized from extended eigenstates and
find its analytical expression for both commensurate and incommensurate modulation wavelengths, thus enlarging the library of models possessing analytical expressions for the ME. Our results extend previous results for the theory of fermionic topological insulators and localization theory in quantum matter to the classical realm.
\end{abstract}
\maketitle

\section{Introduction\label{sec:Introduction}}

A real crystalline material always possesses some degree of disorder.
Often, the disorder is of Anderson type \cite{Anderson1958,FiftyAnderson}, where a random 
potential is created by a finite and small concentration of impurities
dispersed in the material. It is well know that in three dimensions,
there exists a critical value of the potential intensity $\Delta$
above which a metal-insulator phase transition occurs \cite{Abrahams1979}. In
one dimension, however, all states are localized for any small but
finite value of $\Delta$. 
With the advent of twisted bilayer graphene
\cite{Gadelha2021} a new kind of disorder became easily accessible in condensed
matter, dubbed quasi-periodic disorder. This comes about because, at
certain twist angles, the electrons see a potential which is incommensurate with the lattice they propagate in, due to the presence of the second twisted layer. However, quasi-periodicity in
electronic systems is not new, having a long tradition going back
to studies by Serge Aubry and Gilles Andr{\'e} \cite{AA_paper_original,Paredes2019}. In their original
paper, Aubry and André proposed a one dimensional tight-binding model where electrons
are subjected to a sinusoidal electrostatic potential incommensurate
with the system's lattice. These two authors showed that a metal-insulator
transition takes place only for a finite value of $\Delta=2t$, where
$t$ is the electronic hopping amplitude. All eigenmodes are either exponentially
localized, if $\Delta>2t$, or extended plane waves, if $\Delta<2t$.
This result is at odds with the case of Anderson localization in one-dimensional
systems. 

Together with metal-insulator transitions due to disorder, topological
phase transitions in one and more dimensions \cite{Moore2011}
have also become a field of intense research. For example, robust quantum-state transfer in superconducting qubit chains via topologically protected edge states  \citep{Mei2018} was experimentally realized and quantum-information processing via chiral Majorana edge modes in Kitaev materials \citep{Timoshuk2023} was proposed.
In any spatial dimension, and in the absence of disorder, 
such phase transitions are abrupt, as the system suddenly
jumps from a trivial phase,  hosting no edge states, to a topological
one hosting edge states. Thus, in the latter case, an electronic
system supports dissipationless charge transport due to chiral edge
states (for an adiabatic cyclic evolution of the Hamiltonian we have, for example, the Thouless pumping in 1D \citep{Citro2023}) carried by integer or fractional charge excitations \cite{Ady2016}.  A natural question now arises  on
the interplay between topology and quasi-periodic disorder.

The Su-Schrieffer-Heeger (SSH) model is the ideal platform for investigating
the aforementioned  interplay.  In brief, the SSH model is described by the tight-binding Hamiltonian: $t\psi_{j-1}^{B}+v\psi_{j}^{B}=E\psi_{j}^{A}$
\citep{SSH_original}, where $t,v$ are two hopping amplitudes
that connect lattice sites labeled by the index $j$ and the sublattice
index $A/B$. It is the simplest one-dimensional model with topological
features, possessing two topological phases characterized by winding
numbers $\nu=0$ and $\nu=1$, the latter being topologically non-trivial
and supporting edge-states, special eigenstates
that are localized at the boundaries of the system. The existence (or absence) of edge states can also be predicted through the bulk-boundary correspondence \citep{2016short,basu2022condensed}.
Recent studies have shown that random chiral disorder \citep{Shi2021}
and quasi-periodic intracell hopping modulation \citep{Lu2022}
can induce abnormal topological phase transitions on SSH-based systems. Indeed, in a previous study \citep{Antao2024}, we have showcased how quasiperiodicity can induce complex features in the topological description of the SSH model. In such a system, quasiperiodic modulation of intra and inter-cell hoppings can enable the coexistence of topological edge modes linked to 1D and 2D topological invariants.
Additionally, when this model is subjected to quasi-periodicity, nontrivial
localization phenomena, not present in the original Aubry-André (A-A)
model arise, such as the existence of energy dependent mobility edges
(MEs) and critical phases for multiple values of disorder strength
\citep{Tang2023,Roy2021,Lu2022} appear.

As noted above, Aubry and André studied Anderson localization in an
1D tight-binding model whose disorder applied to a periodic lattice
was of sinusoidal form. The model is described by: $t(\psi_{j+1}+\psi_{j-1})+V{\rm cos}(2\pi\beta j+\phi)\psi_{j}=E\psi_{j}$,
where $j$ is the lattice index, $V$ is the strength of the quasi-periodic
potential, $t$ is the hopping amplitude between neighboring sites,
$\phi$ is the phase parameter and $\beta$ is an irrational number
\citep{AA_paper_original}. This model came to be known as the Aubry-André
model, and it is the simplest model concerning quasi-periodicity in which
quasi-disorder can induce abrupt, energy-independent, metal-insulator
transitions: all eigenmodes are either exponentially localized if
$V>2t$ or extended plane waves if $V>2t$. At the condition $V=2t$,
the A-A model is said to posses self duality, meaning that the Hamiltonian
 both in real and momentum spaces has the same identical form upon a dual transformation of the parameters $V$  and $t$.
From there, different quasi-periodic models have been introduced and
understood theoretically \citep{Roy2021,DasSarma1,DasSarma2_ganeshan}
and experimentally realized \citep{AA_experiment1,AA_experiment2,AA_experiment3}. One notable property of the original A-A model is that the boundary separating the extended and localized phases is energy-independent. Thus, no energy-dependent mobility edge (ME)(energy value that marks the boundary
between localized and extended states \citep{Mott,Anderson_MEs}) can exist in this model, unlike in 3D random disordered systems, for example \citep{FiftyAnderson,Evers_anderson_transition}. This is a peculiarity of the Aubry-Andr\'e model, 
since,  in general,  generalized versions of this model possess energy dependent MEs  \citep{Miguel_paper1,Miguel_paper2}. Indeed, it was already shown before that
1D quasi-periodic models can host energy dependent MEs by introducing long-range hopping
amplitudes to the base lattice \citep{DasSarma1} or by applying some
variation of the A-A potential to the Su--Schrieffer--Heeger 
model \citep{Tang2023,Roy2021,Lu2022}.  

In this work, inspired by recent studies of SSH-based models in classical
wave systems, like elastic \citep{Shi2021,mech_SSH1,mech_SSH2,mech_SSH3,mech_SSH4,Chen2018},
photonic \citep{phot_SSH1,phot_SSH1,Lado2023},  acoustic \citep{acoustic_SSH1,acoustic_SSH2,acoustic_SSH3},
and superconducting systems \cite{Qi-Bo2016,Tanatar2019,Jia-Rui2023},
and by exciting localization phenomena induced by quasi-periodicity
\citep{Miguel_paper1,Miguel_paper2,Roy2021}, we investigate a mechanical
version of the SSH model subjected to an A-A intercell spring constant. We focus on the limit where only the intercell elastic spring-constant is altered by the A-A modulation. That is, the intracell elastic constant (between masses $u_{A}^{j},u_{B}^{j}$) remains fixed, whereas the inter
unit cell elastic constant (between $u_{A}^{j+1},u_{B}^{j}$) is allowed to vary. We study the 1D topological
phases of the chiral version of this model numerically and analytically,
in order to understand how variations in the strength, periodicity and relative shift of the A-A potential affect the topological phases and correspondent winding numbers. Then, we make use of the inverse
participation ratio (IPR) and the fractal dimension to study the localization
properties of both chiral and non-chiral version of our model. At
last, the equivalent of mobility edges for our mechanical system is
derived analytically for the non-chiral system based on Avila's global
theory \citep{Avila}.

This paper is organized as follows. In Sec. \ref{sec:model} 
we introduce the model and discuss the topological properties of our classical mechanical model as function of the relationship between springs-constants and the intensity of the intercell spring modulation. In Sec.  \ref{sec:AA-model} we discuss the effect of A-A potential on the winding number phase diagram and on the localization of the eigenstates. Next, we derive an analytical expression of the energy-dependent mobility edge.  Finally, we offer our conclusions and detail some of the calculations in the Appendixes.

\section{Mechanical SSH model}
\label{sec:model}

In this section, we introduce our SSH mechanical model, define the
notation and discuss the model topological properties from a real
space perspective.

\subsection{Model and notation}

\begin{figure}
\includegraphics[scale=0.05]{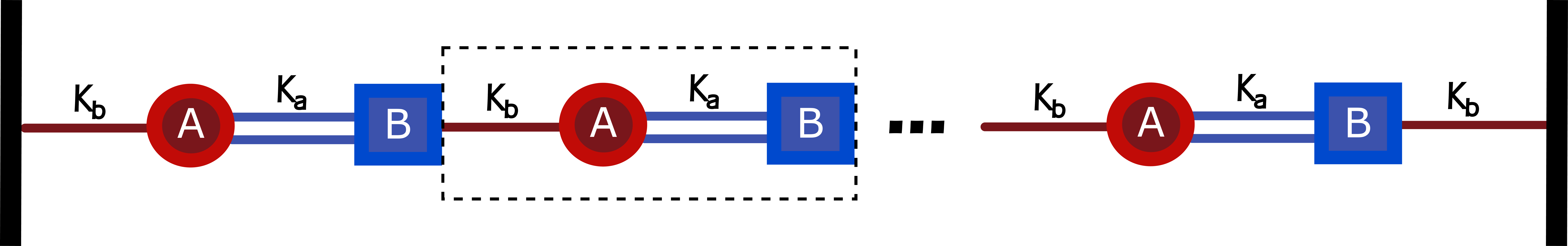}\caption{Depiction of the mechanical SSH with fixed boundary conditions. The
dashed rectangles contains one unit cell.}
\end{figure}

The system is a finite 1D chain subjected to fixed boundary conditions
(FBC), composed of $N$ unit cells with two equal masses $m$ each.
Each mass carries with itself two labels: the sublattice index $\alpha=\{A,B\}$
and the cell index $j\in[1,N]$ . Additionally, two spring constants
are defined: the intracell spring $K_{a}$ and the intercell spring
$K_{b}$. These spring constants correspond to the hopping amplitudes
in the original electronic SSH model and, as will be shown later,
tuning their values can lead to different topological phases in the
mechanical model. The equations of motion obeyed by a mass of the
type $A$ and a mass of the type $B$ are obtained from the Euler-Lagrange
method and can be represented in matrix form as
\begin{equation}
\mathbf{\ddot{U}}(t)=-\mathbf{\mathbb{M}}U(t)
\end{equation}
where $U(t)=[u_{A}^{1},u_{B}^{1},u_{A}^{2},u_{B}^{2},...,u_{A}^{N},u_{B}^{N}]$
is the displacement vector, and $u_{\alpha}^{j}(t)$ is the displacement
of the mass $j$ with sublattice label $\alpha=A/B$. Since the normal
modes have by definition a well defined frequency, we can separate
the temporal and spacial parts. This allows us to write $\ddot{u}_{\alpha}^{j}$
in terms of $u_{\alpha}^{j}$ as follows

\begin{equation}
u_{\alpha}^{j}(t)=C_{\alpha,j}e^{-i\omega t}\rightarrow\ddot{u}_{\alpha}^{j}(t)=-\omega^{2}u_{\alpha}^{j}(t)
\end{equation}
so the equations of motion reduce to the following eigenvalue problem

\begin{equation}
\mathbb{M}U(t)=m\omega^{2}U(t)
\end{equation}
where $\mathbb{M}$ is the dynamical matrix of dimension $2N\times2N$.
It has the form

\begin{equation}
\mathbf{\mathbb{M}}=\left[\begin{array}{ccccc}
K_{a}+K_{b} & -K_{a} & 0 & ... & 0\\
-K_{a} & K_{a}+K_{b} & -K_{b} & ... & 0\\
0 & -K_{b} & K_{a}+K_{b} & ... & 0\\
... & ... & ... & ... & -K_{a}\\
0 & 0 & 0 & ... & K_{a}+K_{b}
\end{array}\right]_{2N\times2N}
\end{equation}
Differently from the original SSH model Hamiltonian, the mechanical
spring-mass dynamical matrix $\mathbb{M}$ has elements in its diagonal.
However, since all diagonal elements are equal, the system still possesses
chiral symmetry, as the diagonal elements represent only a shift in
the eigenvalue spectrum of the form $m\omega_{0}^{2}=K_{a}+K_{b}$.
Then, we can define the matrix $\mathbb{M}'=(\mathbb{M}-m\omega_{0}^{2}\mathbb{I})$
that obeys an anti-commutation relation with the chiral operator
$\varGamma$, and whose eigenvalue spectrum is centered aroud $m\omega^{2}=0$
\[
\varGamma(\mathbb{M}-m\omega_{0}^{2}\mathbb{I})+(\mathbb{M}-m\omega_{0}^{2}\mathbb{I})\varGamma=0
\]
where $\mathbb{I}$ is the $2N\times2N$ identity matrix, and $\Gamma$
is the chiral operator, whose matrix representation for the FBC spring-mass
system is
\begin{equation}
\varGamma=\left[\begin{array}{ccccc}
1 & 0 & 0 & 0 & ...\\
0 & -1 & 0 & 0 & ...\\
0 & 0 & 1 & 0 & ...\\
0 & 0 & 0 & -1 & ...\\
\vdots & \vdots & \vdots & \vdots & \ddots
\end{array}\right]_{2N\times2N}
\end{equation}

\subsection{Topological characterization in real space}

For very large systems, translational symmetry is exact in the bulk (this is not the case near the system's terminations).
Due to this property, it is possible to define topological markers
in real space \citep{Shi2021,Meier2018,Mondragon2014}, which
allows to discriminate between topologically trivial and topologically
non-trivial phases in non-periodic systems. In the present work we
only consider systems that do not possess translational symmetry. The usual winding number
computation in momentum space for periodic systems \citep{Chen2018} must
therefore be handled in real space. For this purpose we use
the concept of local topological marker (LTM) as defined in \citep{Meier2018}.
A LTM has a local value for each unit cell and, when averaged away
from the border, it converges to the winding number of the periodic
system. The advantage of the LTM is that it can also deal with disordered
systems \citep{Shi2021,Meier2018,Mondragon2014}. To compute the LTMs for a chain with $N$ unit cells, a $2N\times2N$
matrix $\mathbb{U}=\left[U_{1},U_{2},U_{3}...,U_{2N}\right]$ is constructed,
whose columns are normalized eigenvectors of the system in ascending
order (the first column is then the eigenvector corresponding to the
smallest eigenvalue). $\mathbb{U}$ can be divided in two: $\mathbb{U_{-}}=\left[U_{1},U_{2},U_{3}...,U_{N}\right]$
containing the eigenvectors bellow band gap at zero energy and $\mathbb{U}_{+}=\left[U_{N+1},U_{N+2},U_{N+3}...,U_{2N}\right]$
containing the eigenvectors above the band gap at zero energy. With $\mathbb{U}_{+}$
and $\mathbb{U}_{-}$ we can construct the projectors of the bands
above and bellow the gap: $\mathbb{P}_{+}=\mathbb{U}_{+}\mathbb{U}_{+}^{T}$
and $\mathbb{P}_{-}=\mathbb{U}_{-}\mathbb{U}_{-}^{T}$ respectively.
The ``flat-band Hamiltonian'' \citep{Mondragon2014} is defined
as $\mathbb{Q}=\mathbb{P}_{+}-\mathbb{P}_{-}$, which can be decomposed
as $\mathbb{Q}=\mathbb{Q}_{AB}+\mathbb{Q}_{BA}=\Gamma_{A}\mathbb{Q}\Gamma_{B}+\Gamma_{B}\mathbb{Q}\Gamma_{A}$,
where
\begin{equation}
\begin{array}{c}
\varGamma_{A}=\left[\begin{array}{ccccc}
1 & 0 & 0 & 0 & ...\\
0 & 0 & 0 & 0 & ...\\
0 & 0 & 1 & 0 & ...\\
0 & 0 & 0 & 0 & ...\\
\vdots & \vdots & \vdots & \vdots & \ddots
\end{array}\right]_{2N\times2N}\\
\varGamma_{B}=\left[\begin{array}{ccccc}
0 & 0 & 0 & 0 & ...\\
0 & 1 & 0 & 0 & ...\\
0 & 0 & 0 & 0 & ...\\
0 & 0 & 0 & 1 & ...\\
\vdots & \vdots & \vdots & \vdots & \ddots
\end{array}\right]_{2N\times2N}
\end{array},
\end{equation}
are the sub-lattice projectors and $\Gamma=\Gamma_{A}-\Gamma_{B}$
is the chiral operator. Finally, the definition of LTM according to
Ref. \citep{Meier2018} is

\small
\begin{equation}
\nu(j)=\frac{1}{2}\sum_{\alpha=A,B}\left\{ \left(\mathbb{Q}_{BA}\left[\mathbb{X},\mathbb{Q}_{AB}\right]\right)_{j\alpha,j\alpha}+\left(\mathbb{Q}_{AB}\left[\mathbb{Q}_{BA},\mathbb{X}\right]\right)_{j\alpha,j\alpha}\right\} 
\end{equation}

\normalsize

where $\mathbb{X}$ is the position operator with dimensions ${\rm dim}(\mathbb{X})=2N\times2N$,
in which the masses are mapped to the position of their unit cell
counting from the center of the system: $\mathbb{X}={\rm diag}[-N,-N,-(N-1),-(N-1),\ldots,N-1,N-1]$,
where the symbol ${\rm diag}[\ldots]$ stands for a diagonal matrix.

\begin{figure}
\raggedright{}%
\begin{minipage}[t]{0.40\linewidth}%
\begin{center}
\includegraphics[scale=0.20]{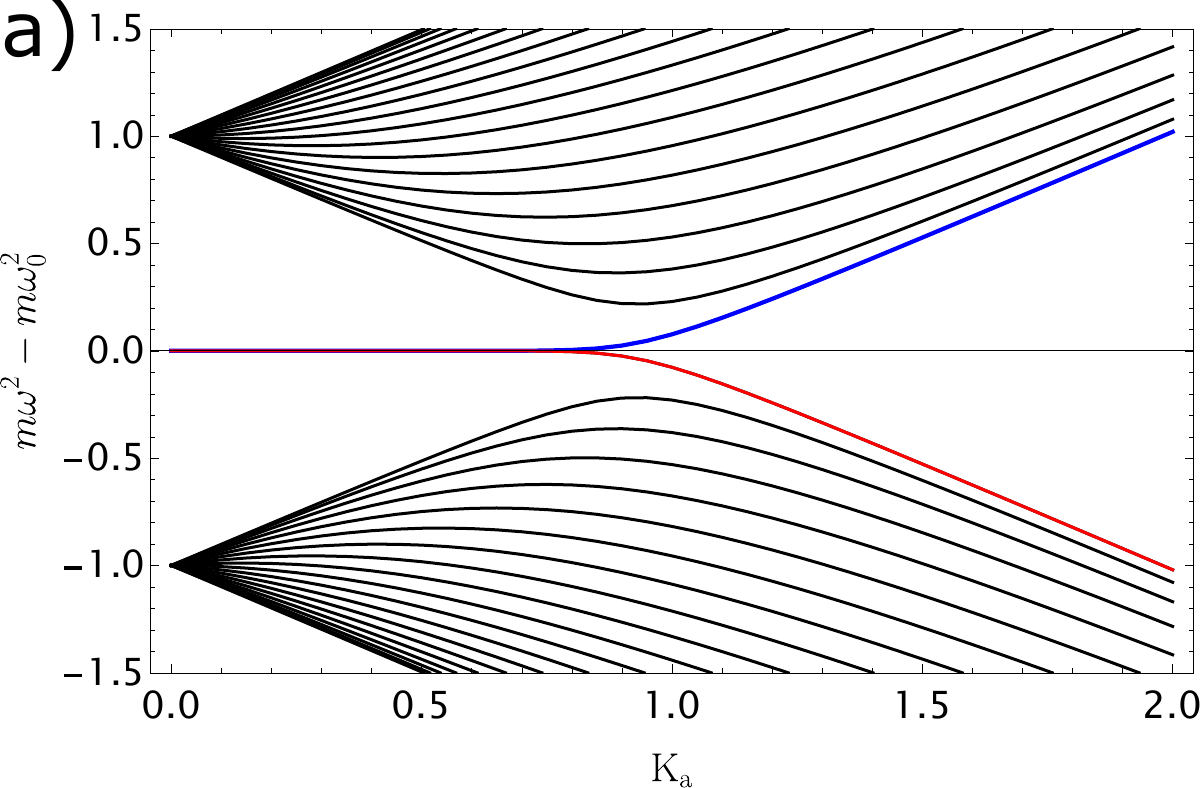}
\par\end{center}%
\end{minipage}\hspace{0.1\linewidth}%
\begin{minipage}[t]{0.40\linewidth}%
\begin{center}
\includegraphics[scale=0.20]{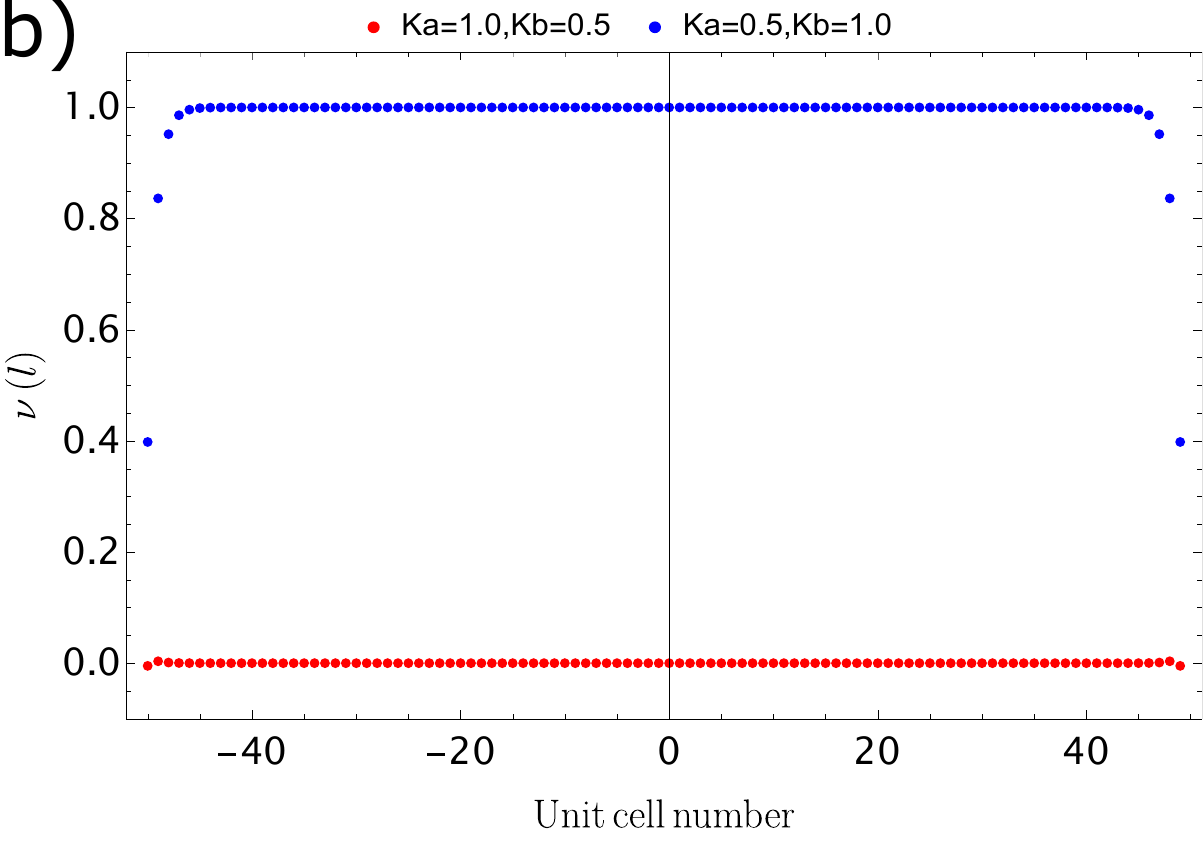}
\par\end{center}%
\end{minipage}\hspace{0.02\textwidth}%
\begin{minipage}[t]{0.40\linewidth}%
\begin{center}
\includegraphics[scale=0.20]{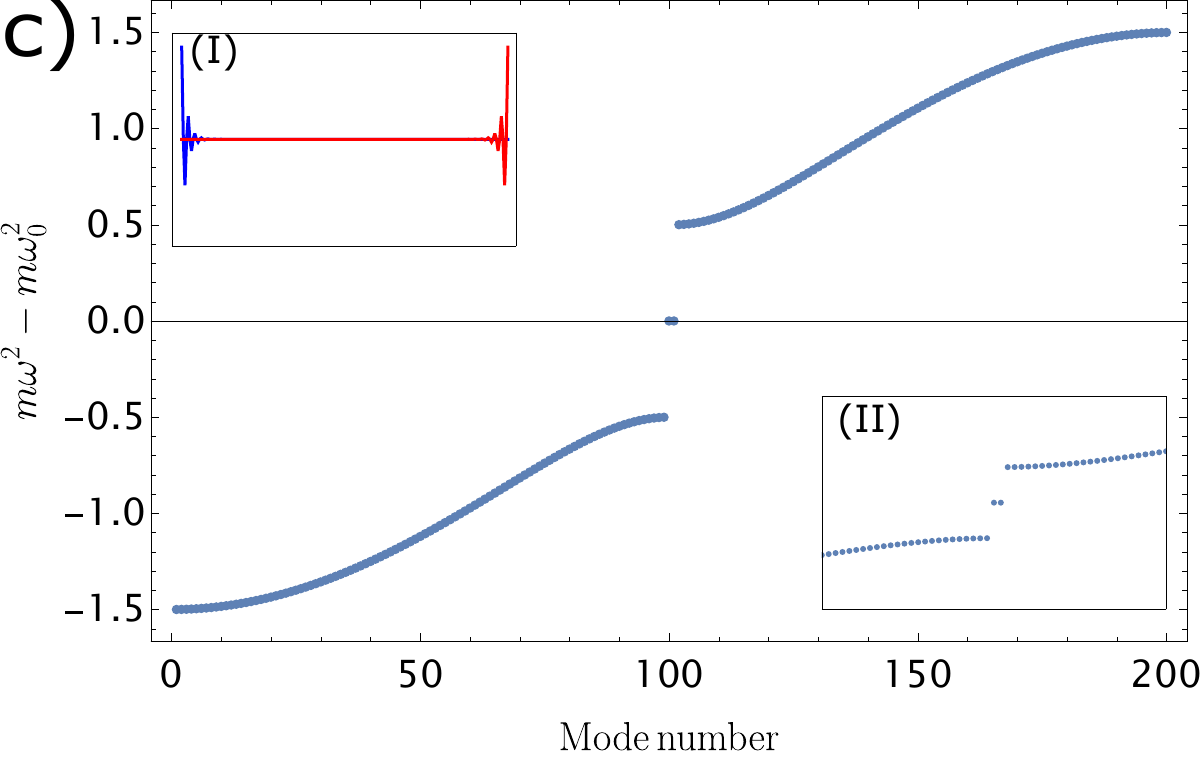}
\par\end{center}%
\end{minipage}\hspace{0.1\linewidth}%
\begin{minipage}[t]{0.40\linewidth}%
\begin{center}
\includegraphics[scale=0.20]{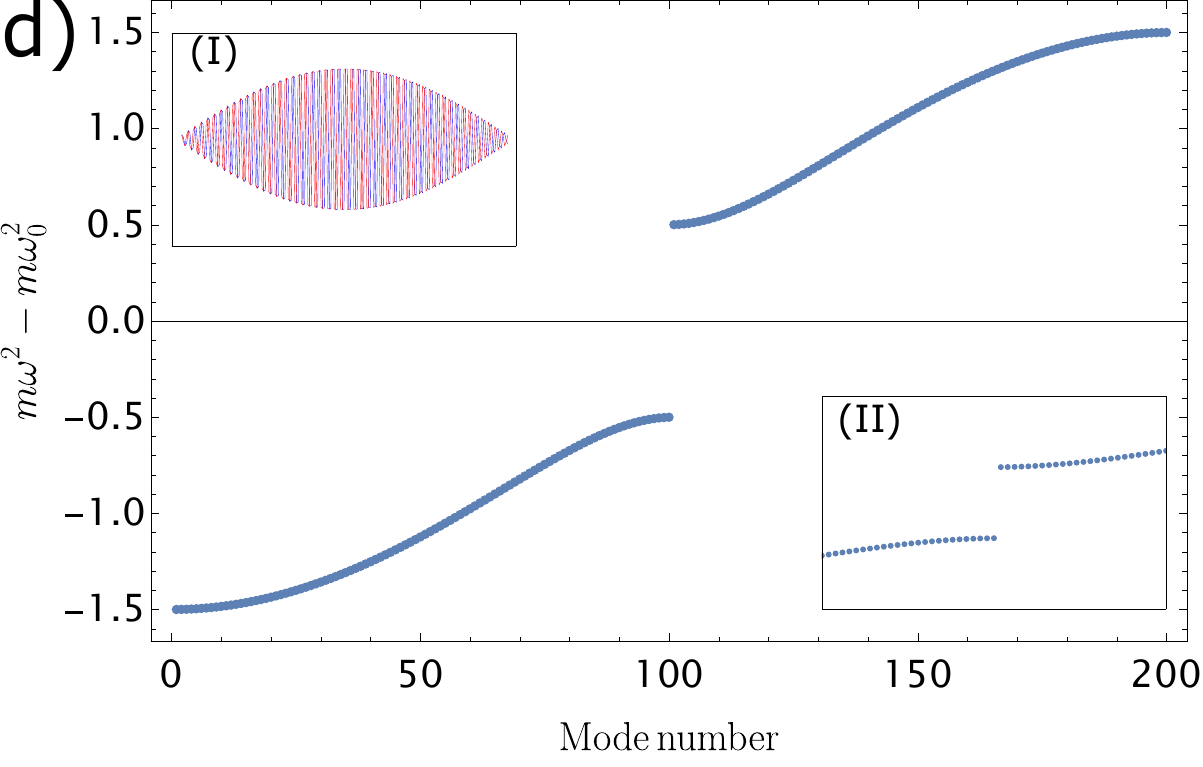}
\par\end{center}%
\end{minipage}\caption{$\mathbf{a)}$ Evolution of the eigenvalue spectrum of $\mathbb{M}'=(\mathbb{M}-m\omega_{0}^{2}\mathbb{I})$
for $K_{b}=1$, $m=1$. The edge states, depicted in red and blue,
are zero-energy states for $K_{a}<K_{b}$ and become bulk states for
$K_{b}<K_{a}$; $\mathbf{b)}$ LTMs calculated for $K_{a}=1,K_{b}=0.5$
(red curve) and $K_{a}=0.5,K_{b}=1$ (blue curve). For unit cells
distant from the borders, the LTMs $\nu(j)$ converge to the winding
number $\nu=0,1$; $\mathbf{c)}$ Eigenvalue spectrum of $\mathbb{M}'$
for $K_{b}=1,K_{a}=0.5$. $\mathbf{c\,I)}$ Eigenmodes number 500
and 501 are localized at the borders and are thus labeled edge-states;
$\mathbf{\mathbf{c\,II)}}$ Edge-states are degenerate and are present
in the middle of the gap.$\mathbf{d)}$Eigenvalue spectrum of $\mathbb{M}'$
for $K_{b}=0.5,K_{a}=1$. $\mathbf{d\,I)}$ Eigenmodes number 500
and 501 are now delocalized; $\mathbf{\mathbf{d\,II)}}$ For this
configuration there are no edge-states. Instead, these eigenmodes
are extended states.}
\label{fig:Energy-spectrum-NN-OBC-edge-states}
\end{figure}

The sub-indexes $jA,jB$ indicate the entries for a mass of type $A$
or $B$ of the $j^{th}$ unit cell in the matrix. The average of the
LTM $\nu(j)$ over the most central cells results in the quantity
$\left\langle \nu\right\rangle $ that converges to the winding number
$\nu$ of the system which, in this model, can have two values: $\nu=0,1$,
characterizing the topologically trivial and non-trivial phases, respectively.
Each phase corresponds to a specific relationship between $K_{a}$
and $K_{b}$ as we show in Fig. \ref{fig:Energy-spectrum-NN-OBC-edge-states}
(a), (b). Additionally, in the topological phase with $\nu=1$, the
system possesses edge-states (or rather, end-states). They
are not present if $\nu=0$ or if we have periodic boundary conditions.
This property is depicted in Fig. \ref{fig:Energy-spectrum-NN-OBC-edge-states}
(c), (d), where the energy spectrum and the shape of the edge-states,
if there are any, can be seen side by side. 

\section{Mechanical model with intercell Aubry-André spring-constants}
\label{sec:AA-model}

In this section we discuss the topological properties, localization and the energy
dependent mobility edges of our mechanical SSH model.

\subsection{Equations of motion}

\begin{figure}
\begin{centering}
\includegraphics[scale=0.05]{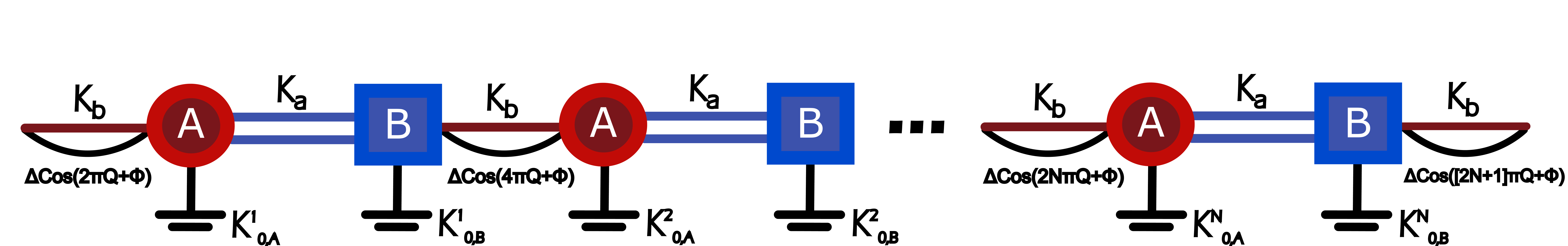}
\par\end{centering}
\caption{Depiction of the mechanical SSH subjected to an Aubry-André modulation
on the intercell spring-constants. In order to preserve chiral symmetry,
a local spring is added in each mass. It depends on the mass type
and on the unit-cell it is in.}

\end{figure}

We begin by studying a nearest-neighbor mechanical model under
influence of an Aubry-André modulation. This modulation is characterized
by a sinusoidal term whose wavelength is incommensurate with the lattice,
due to the irrational number $Q$, chosen here to be the inverse of
the golden ratio $\tau^{-1}=\frac{2}{1+\sqrt{5}}$. Here, we are
interested in studying how the induced quasi-disorder imposed by the
A-A potential affects the topological properties of the system, using
the local topological marker formalism discussed above. In our model, the A-A
is manifested as a perturbation on the springs values, rather than an additional
local spring attached to each mass (this would characterize an on-site
A-A potential). This means that the values of the spring constants
change due to the A-A term. In our model, we apply the Aubry-André
modulation only to intercell springs constants, $K_{b}$, that connects the masses $u_{A}^{j+1} and u_{B}^{j}$. Thus, the equations of motion
for an unit cell $j$ are as follows (time dependence implicit):

\begin{widetext}
\begin{equation}
\begin{array}{c}
m\ddot{u}_{A}^{j}=-(K_{a}+K_{b}+\Delta\mathrm{cos}(2\pi Qj+\phi))u_{A}^{j}+K_{a}u_{B}^{j}+(K_{b}+\Delta\mathrm{cos}(2\pi Qj+\phi))u_{B}^{j-1}\\
\\
\begin{array}{c}
m\ddot{u}_{B}^{j}=-(K_{a}+K_{b}+\Delta\mathrm{cos}(2\pi Q[j+1]+\phi))u_{B}^{j}+K_{a}u_{A}^{j}+(K_{b}+\Delta\mathrm{cos}(2\pi Q[j+1]+\phi))u_{A}^{j+1}\end{array}
\end{array}\label{eq_motion_AA}
\end{equation}
where, $\Delta$ is the A-A amplitude and $j\in[1,N]$ is the unit
cell index. A perturbation on the springs also introduces a term in
the diagonal (this does not happen in the original SSH model), implying
that an off-diagonal A-A potential changes both diagonal and off-diagonal
terms, breaking chiral symmetry. To avoid this, we add local springs
$K_{0,\alpha}^{j}$ to each mass
\begin{equation}
\begin{array}{c}
m\ddot{u}_{A}^{j}=-(K_{a}+K_{b}+\Delta\mathrm{cos}(2\pi Qj+\phi)+K_{0,A}^{j})u_{A}^{j}+K_{a}u_{B}^{j}+(K_{b}+\Delta\mathrm{cos}(2\pi Qj+\phi))u_{B}^{j-1}\\
\begin{array}{c}
\end{array}\\
m\ddot{u}_{B}^{j}=-(K_{a}+K_{b}+\Delta\mathrm{cos}(2\pi Q[j+1]+\phi)+K_{0,B}^{j})u_{B}^{j}+K_{a}u_{A}^{j}+(K_{b}+\Delta\mathrm{cos}(2\pi Q[j+1]+\phi))u_{A}^{j+1}
\end{array},
\end{equation}
\end{widetext}with $K_{0,A}^{j}=-\Delta\mathrm{cos}(2\pi Qj+\phi)$
and $K_{0,B}^{j}=-\Delta\mathrm{cos}(2\pi Q[j+1]+\phi)$. 

This choice of $K_{0,A}^{j}$ and $K_{0,B}^{j}$ preserves the chirality
of the dynamical matrix and allows the computation of a quantized
topological marker $\langle \nu \rangle.$ Assuming normal modes with well defined
frequency, $\ddot{u}_{\alpha}^{j}(t)=-\omega^{2}u_{\alpha}^{j}(t)$
and Eq. (\ref{eq_motion_AA}) becomes

\begin{widetext}
\begin{equation}
\begin{array}{c}
m\omega^{2}u_{A}^{j}=(K_{a}+K_{b})u_{A}^{j}-K_{a}u_{B}^{j}-(K_{b}+\Delta\mathrm{cos}(2\pi Qj+\phi))u_{B}^{j-1}\\
\\
\begin{array}{c}
m\omega^{2}u_{B}^{j}=(K_{a}+K_{b})u_{B}^{j}-K_{a}u_{A}^{j}-(K_{b}+\Delta\mathrm{cos}(2\pi Q[j+1]+\phi))u_{A}^{j+1}\end{array}
\end{array}\label{eq_motion_AA_chiral_version}
\end{equation}

or in matrix form
\begin{equation}
\mathbf{\mathbb{M}}=\left[\begin{array}{ccccc}
K_{a}+K_{b} & -K_{a} & 0 & ... & -(K_{b}+\Delta\mathrm{cos}(2\pi Q+\phi))\\
-K_{a} & K_{a}+K_{b} & -(K_{b}+\Delta\mathrm{cos}(4\pi Q+\phi)) & ... & 0\\
0 & -(K_{b}+\Delta\mathrm{cos}(4\pi Q+\phi)) & K_{a}+K_{b} & ... & 0\\
... & ... & ... & ... & -K_{a}\\
-(K_{b}+\Delta\mathrm{cos}(2\pi Q+\phi)) & 0 & 0 & ... & K_{a}+K_{b}
\end{array}\right]_{2N\times2N}
\end{equation}

\end{widetext}

One important aspect to note is that true quasi-periodicity can only
be achieved in infinite systems, in which the quasi-periodic perturbation
never repeats itself. Numerically, one can only approximate true quasi-periodicity
and its properties by working with sufficiently large systems. However, aim to perform finite size scaling analysis and are thus forced to work with different systems' sizes. In this scenario,
it is necessary to mimic an infinite sized chain by imposing PBC in
the original crystal lattice and defining the appropriate rational
approximant, as detailed in Appendix A, on the additional A-A potential.
This guarantees that the whole system is composed of only one unit cell, in the sense that the A-A potential
never repeats itself throughout the finite system.

\subsection{Topological characterization in the  incommensurate regime}

\begin{figure}
\includegraphics[width=0.43\textwidth]{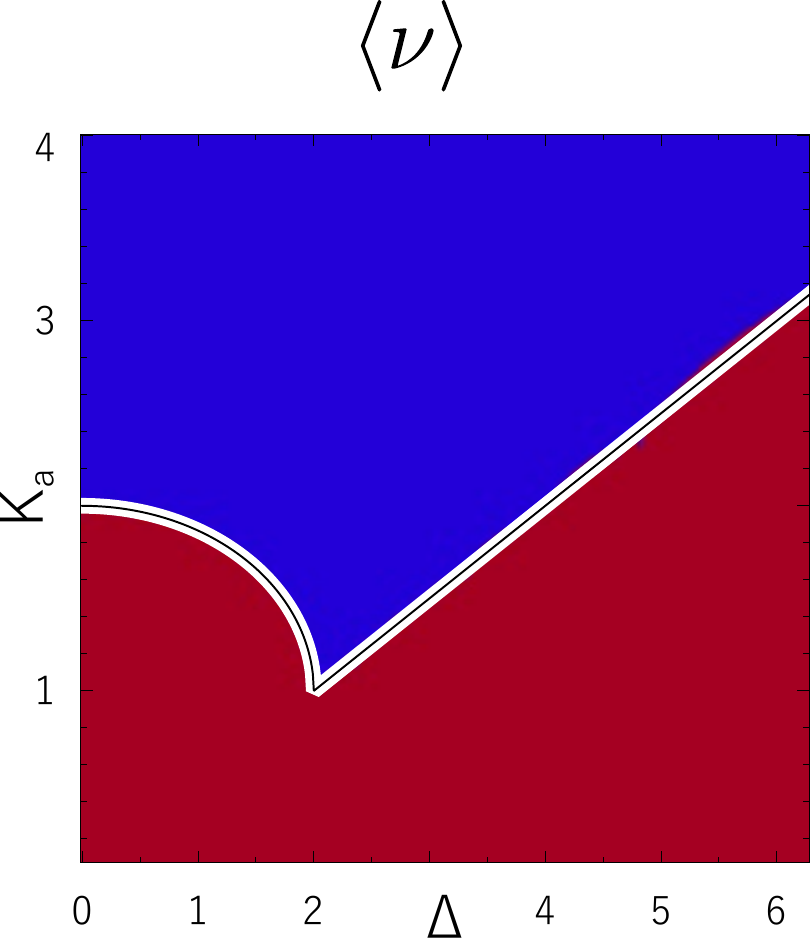}
\caption{Averaged LTMs calculated for the spring-mass system with Aubry-Andr{\'e}
modulation applied on the inter-cell spring-constant. The red region
corresponds to the trivial phase with $\langle \nu \rangle=0$ and the blue region
corresponds to the topological phase $\langle \nu \rangle=1$. Between the two phases, the white stripe separates the two phases and corresponds to a set of points where the averaged LTM did not converge to either 0 or 1 due to the finite system size. The black line is the analytical result for where
the localization length diverges. The agreement between the white and black stripes is a manifestation of the fact that the topological phase transition is accompanied by a divergent localization length.
\label{fig:winding_number}}
\end{figure}

We begin our discussion of the topological phase as a function of
varying $K_{a}$ and $\Delta$, for fixed $K_{b}$. For each combination
of parameters $K_{a}$ and $\Delta$, we calculate the quantity $\langle \nu \rangle$
as described in Section I.B, which converges to the winding number.
Fig. \ref{fig:winding_number} shows the corresponding
phase diagram for $K_{b}=2$, rational approximant $\tilde{Q}=377/610$,
$N=610$ unit cells, $K_{a}\in[0,4]$ and $\Delta\in[0,6]$. As expected
from the original SSH model ($\Delta=0$), the topological phase transition
occurs for $K_{a}=K_{b}$ \citep{2016short}. As $\Delta$
increases from 0 to 2, the trivial topological phase becomes available
for $K_{a}<K_{b}$ due to the quasi-periodic disorder. For $\Delta>K_{b}=2$,
we see that the boundary between the two phases becomes linear and
the phase diagram acquires a re-entrant behavior. In the limit of
very large $\Delta$, the system becomes unable to access the trivial
topological phase, as the intra-cell spring becomes much greater than
the inter-cell spring and the chain converges to a fully dimerized
limit, as if there were no inter-cell springs.

Next, we study the behavior of the localization length. It is well
known \citep{Mondragon2014,Altland2014,Liu2022} that in electronic systems
the topological phase transitions are accompanied by a divergence
of the localization length at the Fermi level. This coincides with
the closing of the band gap. To confirm that this general statement
is also valid for our mechanical model, the localization length $\Lambda$
was calculated analytically and verified to diverge in the region
between topological phases, where the gap closes. In order to simplify
the calculation, we center the eigenvalue spectrum in zero by subtracting
$m\omega_{0}^{2}=(K_{a}+K_{b})$ from each eigenvalue, such that the
states in the middle of the spectrum become zero-energy states when
the gap closes, that is $m\omega^{2}-m\omega_{0}^{2}=0$. Then, the
equations of motion for states with zero energy when the gap closes
read
\begin{equation}
\begin{array}{c}
0=K_{a}u_{B}^{j}+(K_{b}+\Delta\mathrm{cos}(2\pi Qj+\phi))u_{B}^{j-1}\\
\\
\begin{array}{c}
0=K_{a}u_{A}^{j}+(K_{b}+\Delta\mathrm{cos}(2\pi Q[j+1]+\phi))u_{A}^{j+1}\end{array}
\end{array}
\end{equation}
Now, each equation only has coefficients of the same sub-lattice.
The equation with sub-lattice $A$ coefficients, for example, and
can be written as
\begin{equation}
u_{A}^{j+1}=\frac{-K_{a}}{(K_{b}+\Delta\mathrm{cos}(2\pi Q[j+1]+\phi))}u_{A}^{j},
\end{equation}
which leads to a recursive relation
\begin{equation}
u_{A}^{N}=(-1)^{N-1}\prod_{j=1}^{N-1}\frac{K_{a}}{(K_{b}+\Delta\mathrm{cos}(2\pi Q[j+1]+\phi))}u_{A}^{1}
\end{equation}
Now, the relevant Lyapunov exponent $\lambda$, the inverse of the
localization length, can be calculated by \citep{Mondragon2014,Longhi2020,Scales1997}
\begin{equation}
\begin{array}{c}
\gamma=-\lim_{N \to \infty}\frac{1}{N}{\rm Ln}\left|\frac{u_{A}^{N}}{u_{A}^{1}}\right|\\
\\
\begin{array}{c}
=-\lim_{N \to \infty}\frac{1}{N}{\rm Ln}\left|\prod_{j=1}^{N-1}\frac{K_{a}}{(K_{b}+\Delta\mathrm{cos}(2\pi Q[j+1]+\phi))}\right|\end{array}\\
\\
=\lim_{N \to \infty}\frac{1}{N}\left|\sum_{j=1}^{N-1}{\rm Ln}\left|\frac{K_{b}+\Delta\mathrm{cos}(2\pi Q[j+1]+\phi)}{K_{a}}\right|\right|
\end{array}\label{eq:lyapunov_analytical}
\end{equation}
According to Weyl's equidistribution theorem and properties of irrational
rotations \citep{Weyl1916,Choe1993},
a sequence $2\pi Q,4\pi Q,6\pi Q,...$ mod. $2\pi$ is uniformly distributed
in the interval $(-\pi,\pi),$ where $Q$ is an irrational number.
So, the summation in Eq. (\ref{eq:lyapunov_analytical}) can be converted
to an integral over the phase $\varphi\in [0,2\pi]$
\begin{equation}
\begin{array}{c}
\gamma=\frac{1}{2\pi}\intop_{0}^{2\pi}{\rm Ln}\left|\frac{K_{b}+\Delta\cos\varphi}{K_{a}}\right|d\varphi\end{array},
\end{equation}
which can be evaluated as
\begin{equation}
\gamma=\left\{ \begin{array}{c}
{\rm Ln}\frac{K_{b}+\sqrt{K_{b}^{\text{2}}-\Delta^{2}}}{2K_{a}},K_{b}>\Delta\\
\\
\begin{array}{c}
{\rm Ln}\frac{\Delta}{2K_{a}},K_{b}<\Delta\end{array}
\end{array}\right.
\end{equation}

The topological phase transition boundaries can then be computed by considering when the Lyapunov exponent goes to zero, which corresponds to a divergent localization length. Note that with an A-A periodicity incommensurate with that of the lattice, all information about the phase shift $\phi$ which can be initially present in the modulation is washed out by the integration over $\varphi$. This proves that the topological edge modes of such a system will remain insensitive to phase shifts in the modulation, so long as is incommensurate with the lattice. This is not necessarily the case if the modulation is commensurate, and we shall delve into this point a bit more deeply in the following subsection.

Regardless of this subtle point, this analytical result can be verified numerically by computing the
localization length at the central frequency $\omega=\omega_{0}=\sqrt{\frac{K_{a}+K_{b}}{m}}$
using a transfer matrix method \citep{Scales1997}
as a function of $K_{a}$ and $\Delta$. The transfer matrices $T^{i}(\omega)$
are derived from Eq. (\ref{eq_motion_AA_chiral_version}) and have the form

\small
\begin{equation}
\begin{array}{cc}
\left[\begin{array}{c}
u_{i+2}\\
u_{i+1}
\end{array}\right]=T^{i}(\omega)\left[\begin{array}{c}
u_{i+1}\\
u_{i}
\end{array}\right], & T^{i}(\omega)=\left[\begin{array}{cc}
\frac{m\omega^{2}-K_{a}-K_{b}}{K_{i+2}} & -\frac{K_{i+1}}{K_{i+2}}\\
1 & 0
\end{array}\right]\end{array},
\end{equation}

\normalsize

where $i\in[1,2N-2]$ is the index of masses and springs from left
to right, such that $u_{A}^{1}=u_{1}$, $u_{B}^{1}=u_{2}$, $u_{A}^{2}=u_{3}$
and so on. It is important to notice that the first and the last springs
of the system are: $K_{1}=K_{b}+\Delta\mathrm{cos}[2\pi Q+\phi]$
and $K_{2N+1}=K_{b}+\Delta\mathrm{cos}(2\pi Q[2N+1]+\phi)$. So $K_{2}=K_{4}=K_{6}...=K_{2N}=K_{a}.$
Since the transfer matrices of this problem have dimension two, there
exist only two Lyapunov exponents $\lambda_{1}=-\lambda_{2}$ (this
is true if the first and last springs are equal) \citep{Scales1997}.
The localization length is then $\Lambda=\frac{1}{|\lambda_{1}|}=\frac{1}{|\lambda_{2}|}.$
Results for the analytical calculation is shown as a black line in
Fig. \ref{fig:winding_number} for $K_{b}=2$, $m=1$
and $\omega=\omega_{0}=\sqrt{\frac{K_{a}+K_{b}}{m}}$, where it is
clear that the divergence of the localization length for when the
gap closes indeed occurs at the boundary between topological phases
for our model. Thus, this property is not exclusive of electronic models.

\subsection{Topological characterization in the  commensurate regime}

Here we derive the phase transition boundary for topological transition in the commensurate case. In a previous work \cite{Antao2024} we have pointed out the relationship between the 1D topological phase of the model with the number of Dirac cones observed in the 2D spectrum originating from its 2D superspace descripiton.  Here, we build on these previous results by making use of the connection between the mobility edge at zero energy and the critical points where the winding number changes values. In this manner, we completely analytically characterize the 1D topological phase transition boundary, or equivalently the number of Dirac cones in the superspace description of this model. We show that this result reduces to the one obtained in \citep{Antao2024} in the limit of long modulation periodicities.

The starting point of this analysis is, again, Eq. 15. In the incommensurate case, the summation can be converted to an integral due to the properties of rotations by an irrational angle. In the commensurate case this transformation cannot be performed, but nevertheless some simplifications can be implemented. We start from Eq. 15 by replacing $Q=p/q$, where $p$ and $q$ are relatively prime integers

\begin{equation}
\gamma = -\lim_{N\to\infty}\frac{1}{N}\rm{Ln}\left|\prod_{j=1}^{N-1}\frac{K_a}{\left(K_b+\Delta \cos\left(2\pi \frac{p}{q}\left[j+1\right]+\phi\right)\right)}\right|.
\end{equation}

Let us define, for simplicity $\varphi_j^{(q)} = 2\pi p[j+1]/q$. Now, although we cannot convert this expression to an integral such as in the incommensurate case, the periodicity in $q$ nonetheless restrict the number of different terms that appear in the product. Specifically, $j+1$ will yield the same term in the product if $j+1$ mod $ q= j+1$. This means that the product can be restricted to $q$ different terms, and each term will appear $(N-1)/q$ times. The resulting product is

\begin{equation}
\gamma = -\lim_{N\to\infty}\frac{1}{N}\rm{Ln}\left|\prod_{j=1}^{q}\left[\frac{K_a}{\left(K_b+\Delta \cos\left(\varphi_j^{(q)}+\phi\right)\right)}\right]^{(N-1)/q}\right|.
\end{equation}

We can bring a factor $(N-1)/q$ outside the logarithm, and in the large $N$ limit approximate $N-1\approx N$ so that the dependence in $N$ drops from the expression and the limit can also be dropped. We find

\begin{equation}
\gamma = \frac{1}{q}\rm{Ln}\left|\prod_{j=1}^{q}\left[\frac{\left(K_b+\Delta \cos\left(\varphi_j^{(q)}+\phi\right)\right)}{K_a}\right]\right|.
\label{eq:Polynomial_Topological_Boundary}
\end{equation}

Similarly to the incommensurate case, the 1D topological transition boundary can be found by considering the set of parameters where the Lyapunov exponent goes to zero. This is given, in the commensurate case, by the polynomial equation 

\begin{equation}
K_a^q = e^{i \theta_q}\prod_{j=1}^{q} \left[K_b+\Delta \cos\left(\varphi_j^{(q)}+\phi\right)\right]
\label{eq:Polynomial_Topological_Boundary}
\end{equation}

This polynomial equation is of degree $q$ and can be solved for any commensurate periodicity. $e^{i\theta_q}$ is an arbitrary global phase which must be chosen so that all relevant parameters $K_a$, $K_b$  and $\Delta$ are real.  For small periodicity, i.e. in the limit $pN/q \ll 1$, we can send $p/q\to 0$, and recover the transition obtained analytically for the large periodicity limit in \citep{Antao2024}. This is simply the boundary

\begin{equation}
K_a = \pm \left(K_b + \Delta \cos\left(\phi\right)\right)
\label{eq:Polynomial_Topological_Boundary_large}
\end{equation}

\begin{figure}
\includegraphics[scale=0.3]{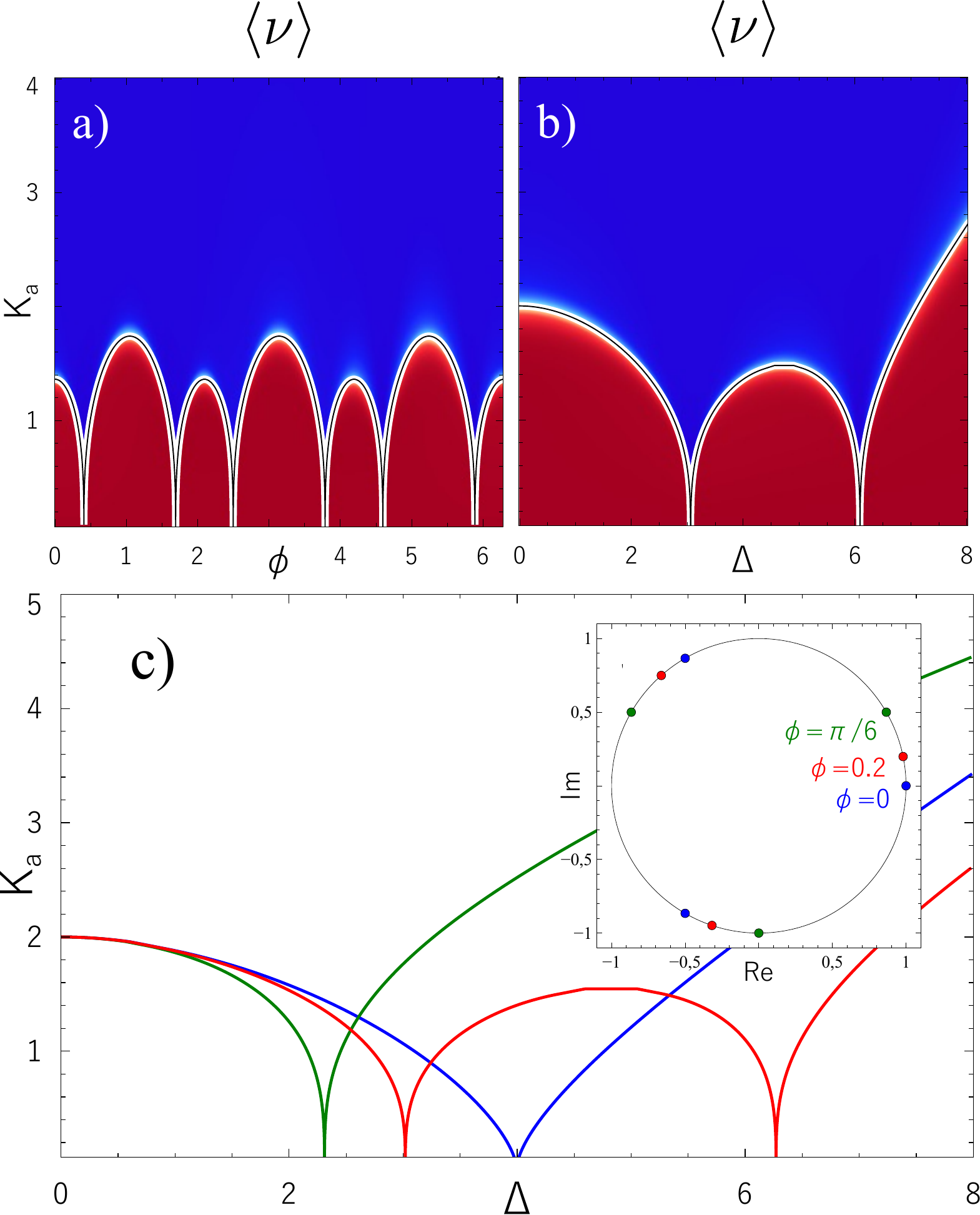}%
\caption{$\mathbf{a)}$ Averaged LTMs calculated for the spring-mass system with Aubry-Andr{\'e}
potential applied on the inter-cell spring-constant in the case of a commensurate modulation of $Q=1/3$ as a function of $\phi$ and $K_a$ and for $K_b=2$ and $\Delta=2.5$.  The blue region
corresponds to the trivial phase with $\langle\nu\rangle=0$ and the red region
to the topological phase $\langle\nu\rangle=1$. The white stripe indicates the region where the topological invariant did not converge to 0 or 1 due to the system's finite size, and the black line to the analytical result for the divergent localization length. $\mathbf{b)}$ Similar phase diagram as a function of $\Delta$ and $K_a$. $\mathbf{c)}$. Different topological transition boundaries as derived analytically from the divergent localization length. The inset shows, for each of the three phases $\phi=0,0.2,\pi/6$, the $q=3$ angles $\varphi_j^{(q)}$ contributing to the product in Eq. \ref{eq:Polynomial_Topological_Boundary}.}
\label{fig:Topological_Commensurate}
\end{figure}

We see that the possibility of manipulation of the 1D topological edge modes becomes very rich in the case of a  commensurate modulation, since an additional dependence on the modulation shift $\phi$ appears, whereas in the incommensurate case it is washed out by the angular integral. Thus, by tuning both the A-A disorder strength as well as this shift, a rich landscape of topological phase diagrams can appear. An example of such a phase diagram as function of  $\Delta$ and $\phi$ is given in Fig. \ref{fig:Topological_Commensurate} a) and b) respectively. In both situations, an initially topologically trivial system can be driven into a topological phase via the A-A disorder term by changing the value of the quasi-disorder strength or merely the shift $\phi$. The fact that the contributions from the product don't wash out the dependence in $\phi$ is showcased in the inset of Fig. \ref{fig:Topological_Commensurate} c), where inspecting the symmetry of the distribution of the angles $\varphi_j^{q}$ makes it clear that changing $\phi$ will result in measurable effects in the topological phase diagram. This is illustrated by plotting the analytical result for phase transition boundary in panel c) as well as directly via the computation of the averaged LTM in panel a).

\subsection{Localization properties, inverse
participation ratio, and finite size scaling analysis}

\begin{figure}
\raggedright{}%
\begin{minipage}[t]{0.45\linewidth}%
\includegraphics[scale=0.15]{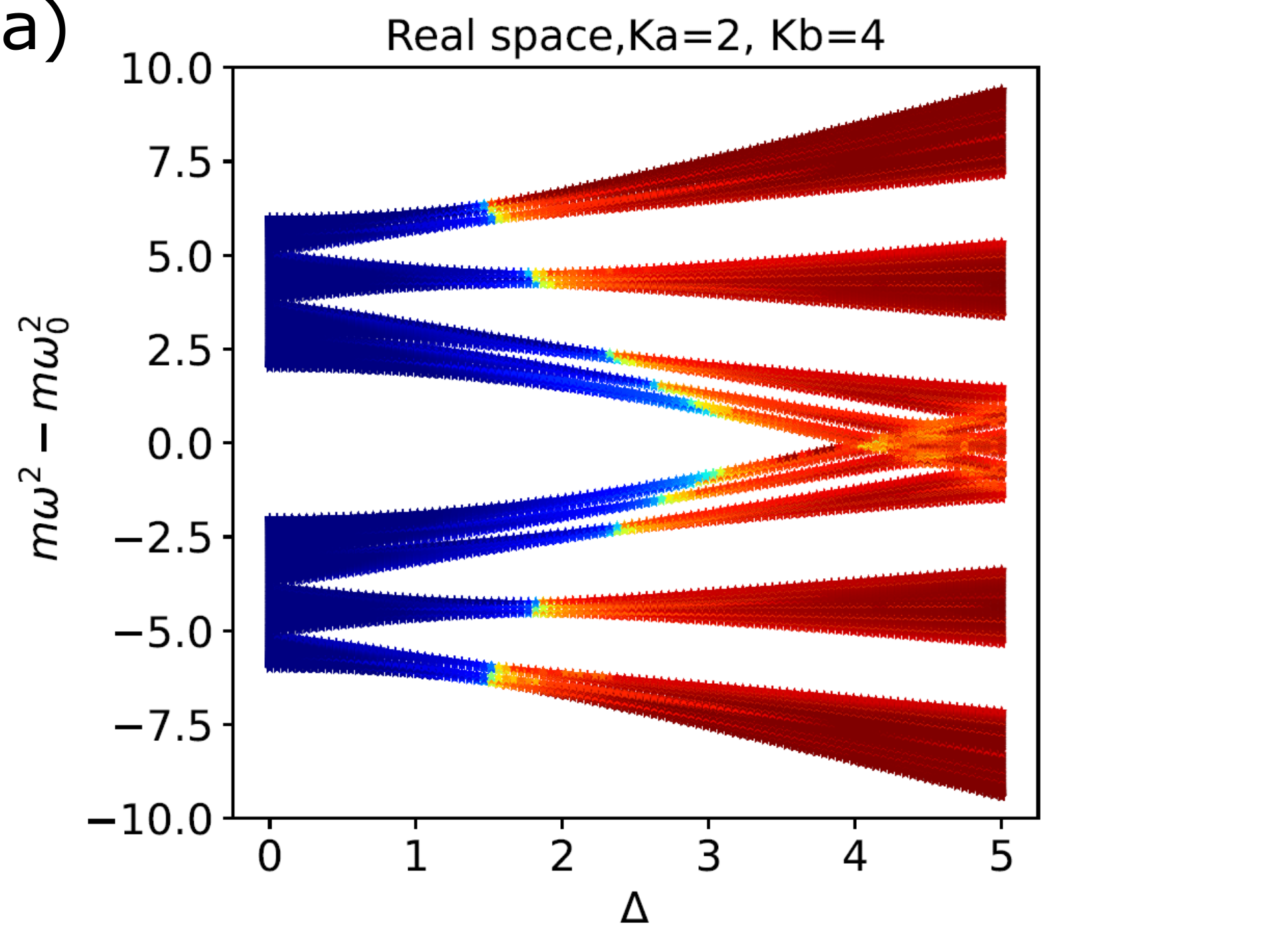}%
\end{minipage}\hspace{0.04\linewidth}%
\begin{minipage}[t]{0.45\linewidth}%
\includegraphics[scale=0.15]{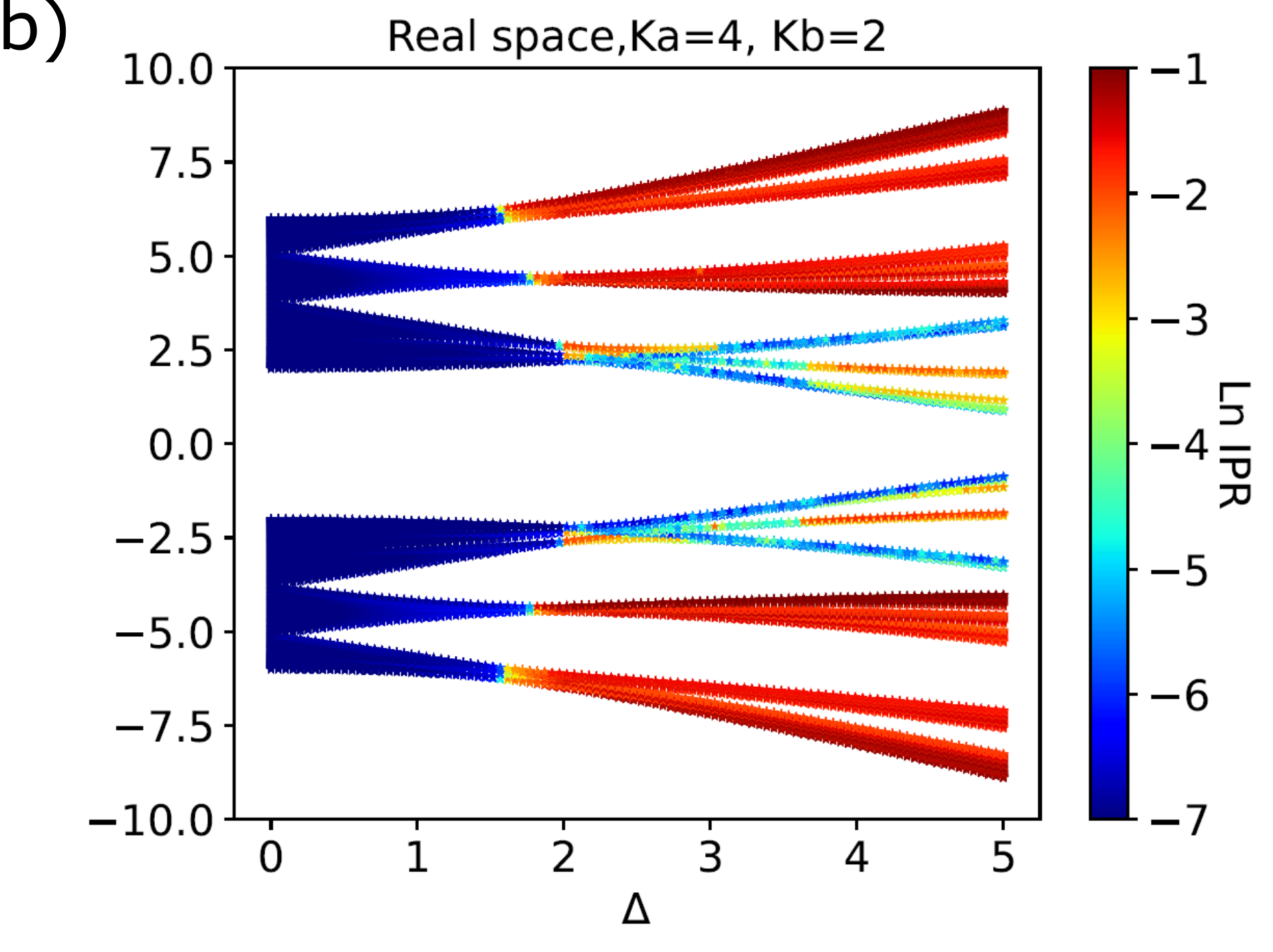}%
\end{minipage}\hspace{0.05\linewidth}%
\begin{minipage}[t]{0.45\linewidth}%
\includegraphics[scale=0.15]{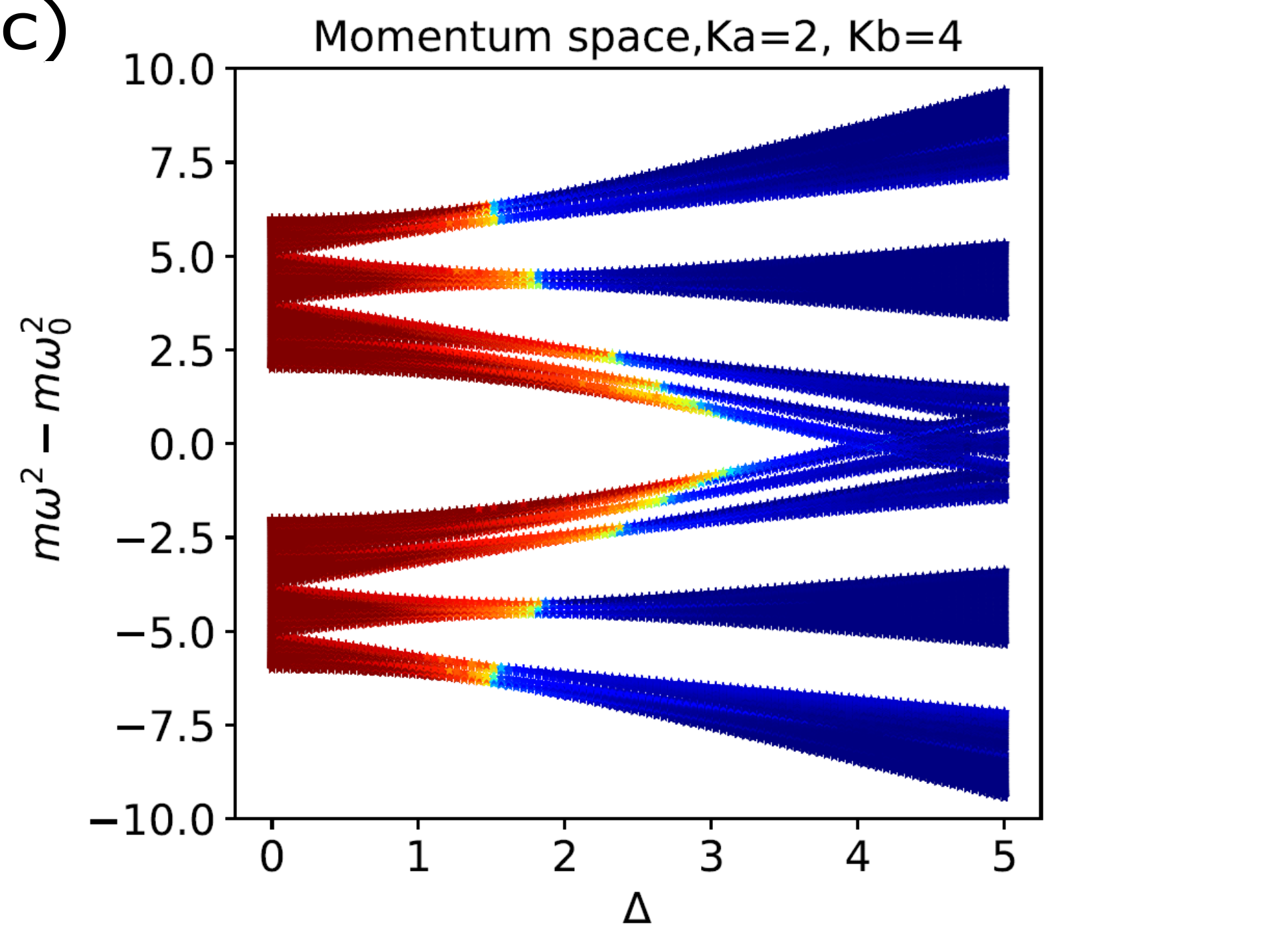}%
\end{minipage}\hspace{0.04\linewidth}%
\begin{minipage}[t]{0.45\linewidth}%
\includegraphics[scale=0.15]{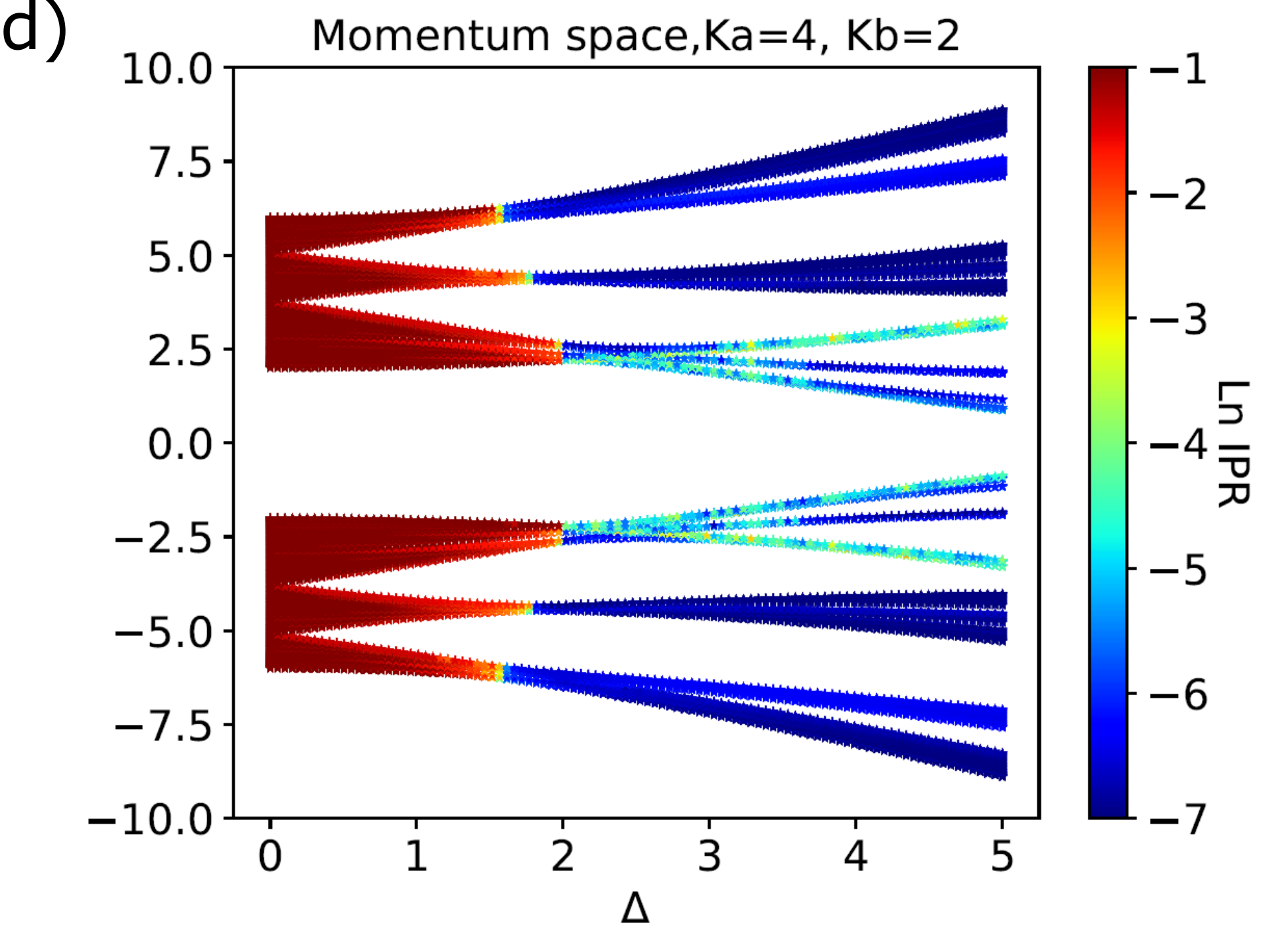}%
\end{minipage}
\caption{Ln(IPR) as a function of the eigenvalues $(m\omega^{2}-m\omega_{0}^{2})$
for $N=987$ unit cells and $\mathbf{a)}$$K_{b}>K_{a}$ in real space;
$\mathbf{b)}$$K_{a}>K_{b}$ in real space; $\mathbf{c)}$$K_{b}>K_{a}$
in momentum space; $\mathbf{d)}$$K_{a}>K_{b}$ in momentum space.
If $K_{b}>K_{a}$, the transition between extended (localized) to
localized (extended) phases in real (momentum) space is sharp, while
for $K_{a}>K_{b}$ there are regions in which the transition is sharp
and regions in which the state IPR does not converge to 1 (0). This could be an indication that the system is in the critical phase. However, in order to verify this statement, it is necessary to perform a finite size scaling analysis. See Figs.  \ref{fig:D(N)}  and \ref{fig:gamma} and corresponding discussion in the main text. \label{fig:painted_IPR_diagrams}}
\end{figure}

Localization properties can also be accessed analysing the inverse
participation ratio (IPR). The state IPR is defined as \citep{IPR_Bell_1972,Thouless1974,Cai2022}:
\begin{equation}
{\rm IPR}(\left|\psi\right\rangle )=\sum_{i=1}^{2N}|c_{i}|^{4}
\end{equation}
given a normalized eigenstate $\left|\psi\right\rangle $ with dimension
$2N$ and components $c_{i}$.
If the eigenmode is an extended state,
IPR $\simeq N^{-1}$. In the opposite scenario, the eigenmode is
perfectly localized and IPR $\simeq N^{0}=1$. Additionally, states
are said to be in the critical phase if they are neither fully localized nor fully extended.  In this case its IPR does not converge to zero as the system
increases in size \citep{Saul1988,Evers_anderson_transition,Roy2021,Cai2022,Callum2024} and a situation occurs where localized and critical states coexist.
The states' IPR can be represented in a graph of the energy eigenvalues as function of 
 the Aubry-André potential strength $\Delta$, with the lines colored by the value of the IPR of the corresponding eigenstate. 
  Such
representation allows us to visualize the transition between extended,
localized,  and critical phases, as shown in Fig. \ref{fig:painted_IPR_diagrams}
for both real and momentum spaces. The transformation to momentum
space is detailed in Appendix B.

\begin{figure}
\begin{minipage}[t]{0.45\linewidth}%
\includegraphics[scale=0.3]{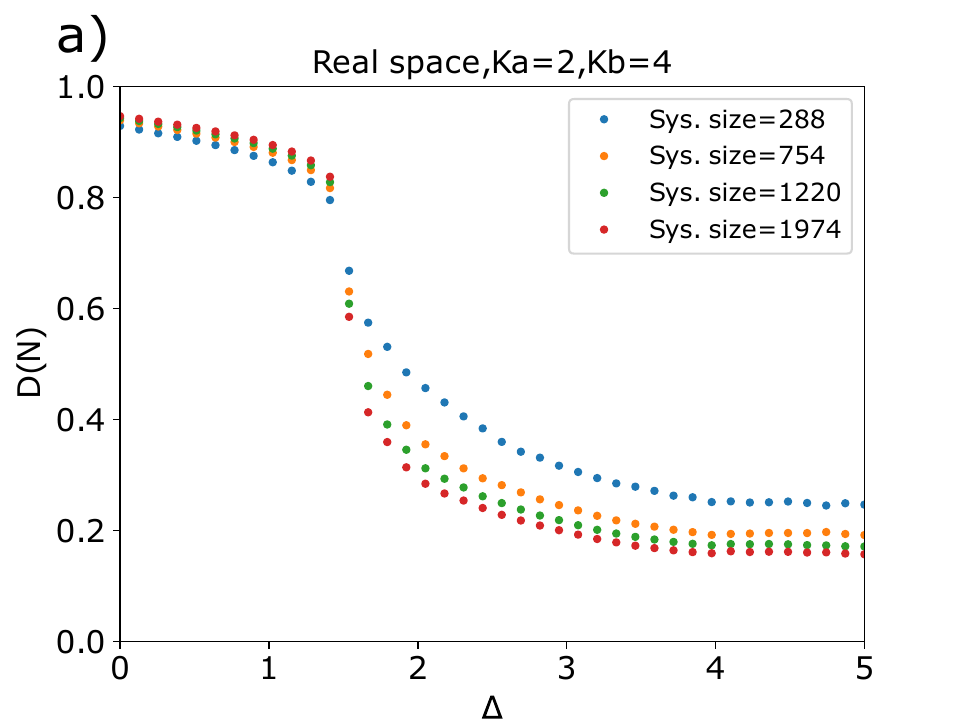}%
\end{minipage}\hspace{0.08\linewidth}%
\begin{minipage}[t]{0.45\linewidth}%
\includegraphics[scale=0.3]{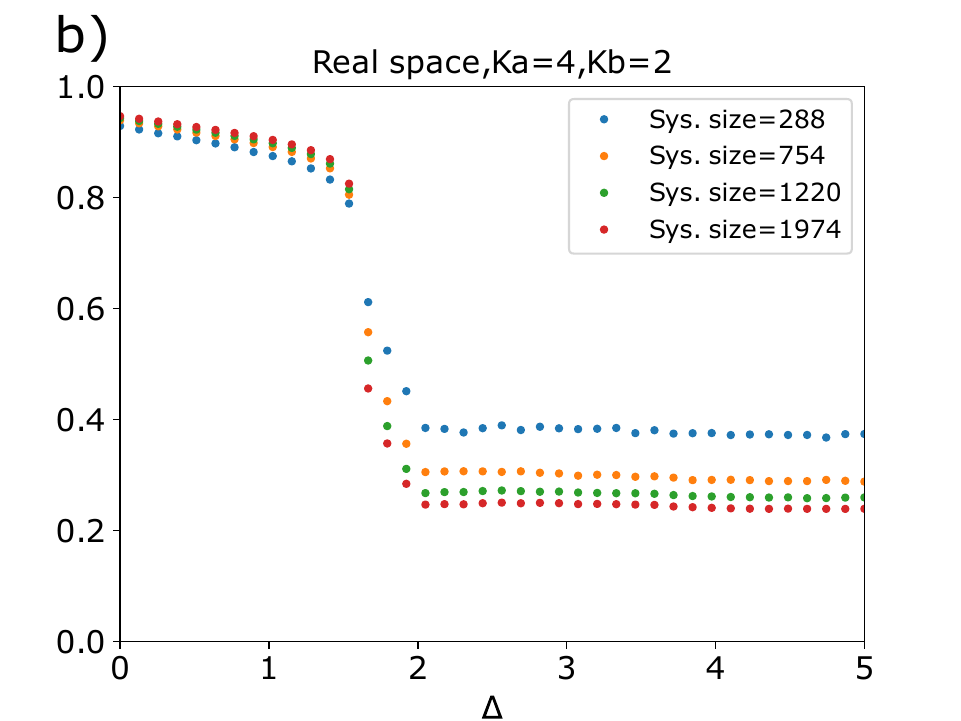}%
\end{minipage}\hspace{0.05\linewidth}%
\begin{minipage}[t]{0.45\linewidth}%
\includegraphics[scale=0.3]{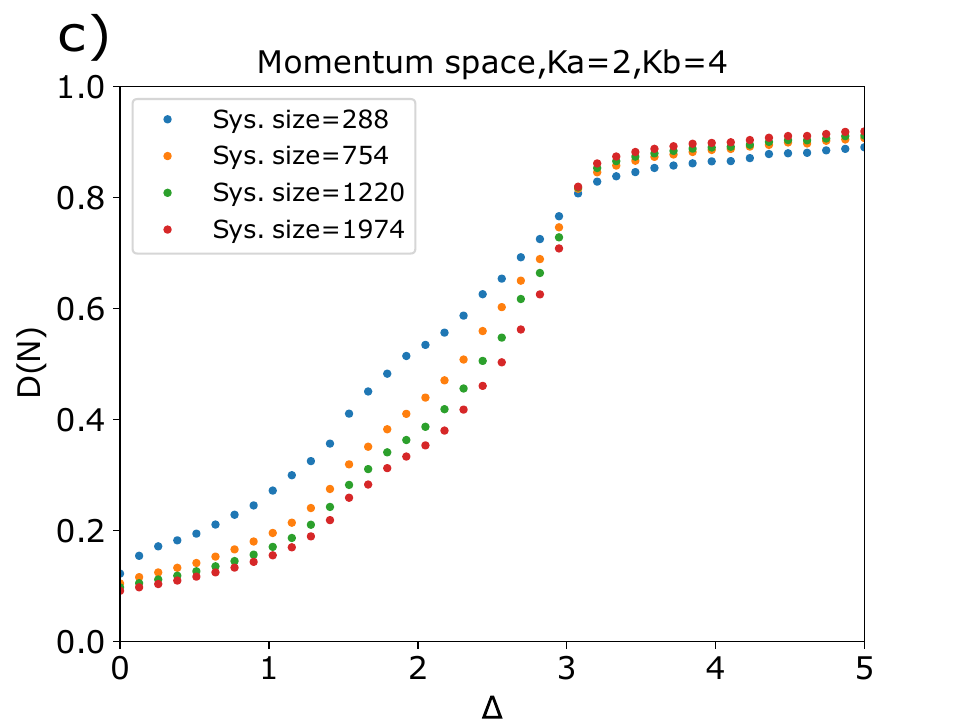}%
\end{minipage}\hspace{0.08\linewidth}%
\begin{minipage}[t]{0.45\linewidth}%
\includegraphics[scale=0.3]{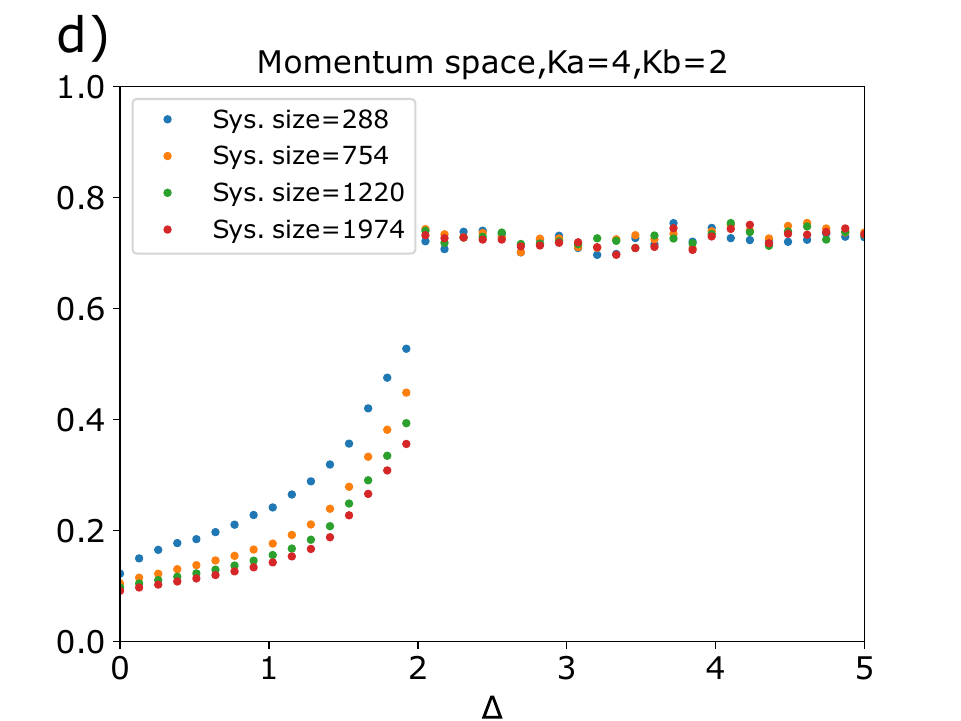}%
\end{minipage}\caption{Fractal dimension $D(N)$ as a function of $\Delta$ and the system
size $2N$ for $\mathbf{a)}$ $K_{b}>K_{a}$ in real space; $\mathbf{b)}$ $K_{a}>K_{b}$
in real space; $\mathbf{c)}$ $K_{b}>K_{a}$ in momentum space; $\mathbf{d)}$ $K_{a}>K_{b}$
in momentum space. In cases $\mathbf{a)}$, $\mathbf{b)}$ the system
begins in an extended phase, since $D(N)$ approaches 1 with increasing
$N$. Then, it transitions rather quickly to a localized phase as
$\Delta$ increases, since $D(N)$ approaches 0 with increasing $N$.
In $\mathbf{c)}$ the system begins in a localized phase and transition
slowly to an extended phase. In $\mathbf{d)}$ the behavior for $\Delta<2$
is similar to case $\mathbf{c)}.$ After this point, however, the
value of $D(N)$ is the same independent of $N.$ This indicates that
the system is indeed in the critical regime. \label{fig:D(N)}}
\end{figure}

When analysing localization, one convenient quantity to compute is the fractal dimension $\Gamma$.   
This quantity is independent of the system size and controls the asymptotic behavior of the system's IPR (defined as the average of all state IPRs and referred here as sIPR) with $N$ according to the scaling law

\begin{equation}
{\rm sIPR}(N) \sim (2N)^{-\Gamma},
\end{equation}
in the limit $N\rightarrow \infty$. One way of accessing the value of $\Gamma$ from the results for finite size systems is via the introduction of the size dependent quantity $D(N)$, that converges to $\Gamma$ in the thermodynamic limit
\begin{equation}
 \lim_{N \to \infty}  D(N)= \Gamma\,, 
\end{equation}
where $D(N)$ is defined as the average fractal dimension
\begin{equation}
D(N)= -\frac{{\rm Ln}({\rm sIPR})}{{\rm Ln}(2N)} ,
\end{equation}
where $2N$
is the size of the system. For the extended phase, $\Gamma = 1$ and for the localized phase, $\Gamma = 0$. If $1 > \Gamma > 0$, the system is said to be critical \citep{Evers_anderson_transition,Lu2022}. Numerically, one can study the behavior of the quantity $D(N)$ that, for increasing $N$, approaches $\Gamma$.  The advantage of computing $D(N)$ over computing only the IPRs or sIPR is
that it gives a clear criteria  for distinguish between critical, extended, and localized phases,  thus making it possible to predict the phase of the system in the thermodynamic limit. We
then should expect that, for a given $\Delta$, $D(N)$ converges to
1 with increasing $N$ if the system is in an extended phase (all eigenstates are extended).   On the
contrary, $D(N)$ converges to $0$ in the localized regime (where a finite fraction of the states are localized). If the system
is in the critical regime, $D(N)$ assumes a finite value between 0 and
1 independent of $N$. For the cases depicted in Fig. \ref{fig:painted_IPR_diagrams}
we compute $D(N)$ as a function of $\Delta$ and different system sizes,
namely $N=144,377,610$ and $987$ unit cells. Results are shown in
Fig. \ref{fig:D(N)}. In the real space, as seen in Fig. \ref{fig:D(N)}(a) and (b), the system seems to be in the extended phase for small $\Delta$, since the greater the system size, the closer $D(N)$ is to 1. At around $\Delta = 1.5$, there is a transitions to the localized phase, hinted by the fact that $D(N)$ approaches 0 for increasing $N$. This means that the critical states in Fig. \ref{fig:painted_IPR_diagrams} (b) will become progressively localized for larger system sizes.  In  momentum space, as seen in Fig. \ref{fig:D(N)}(c) and (d), the system begins localized for small $\Delta$. In the case $K_{a}=2,K_{b}=4$, there is a transition to the extended phase close to $\Delta = 3$. Interestingly, for $K_{a}=4,K_{b}=2$ and $\Delta > 2$, the ratio $D(N)$ oscillates around a constant value between 0 and 1 independent of $N$. This is a strong evidence that the system tends to the critical phase in the thermodynamic limit. 

\begin{figure}
\raggedright{}%
\begin{minipage}[t]{0.45\linewidth}%
\includegraphics[scale=0.3]{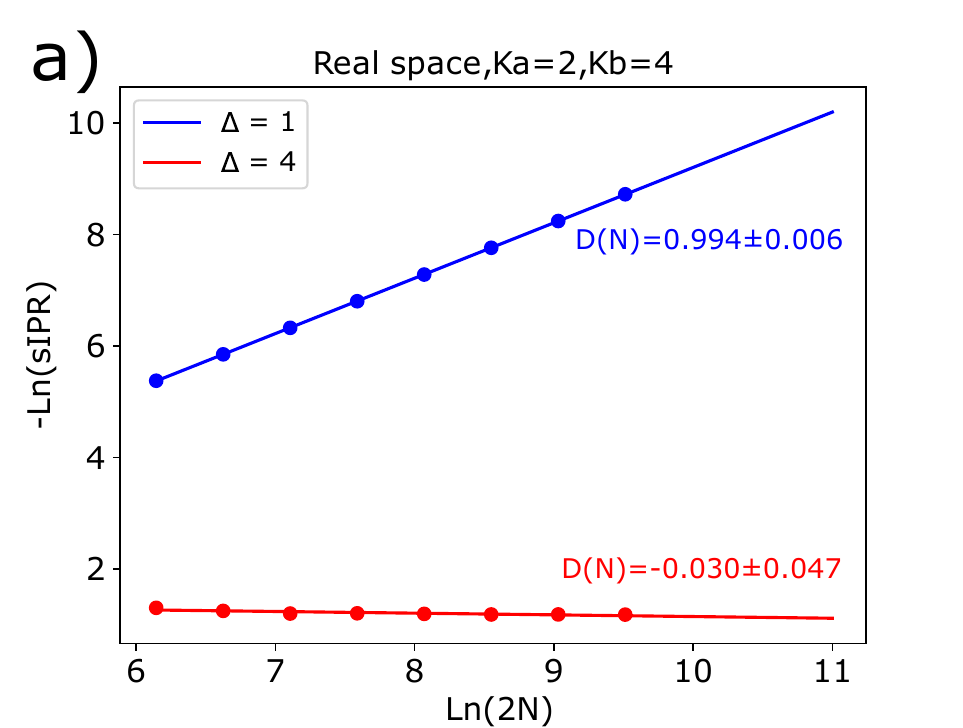}%
\end{minipage}\hspace{0.07\linewidth}%
\begin{minipage}[t]{0.45\linewidth}%
\includegraphics[scale=0.3]{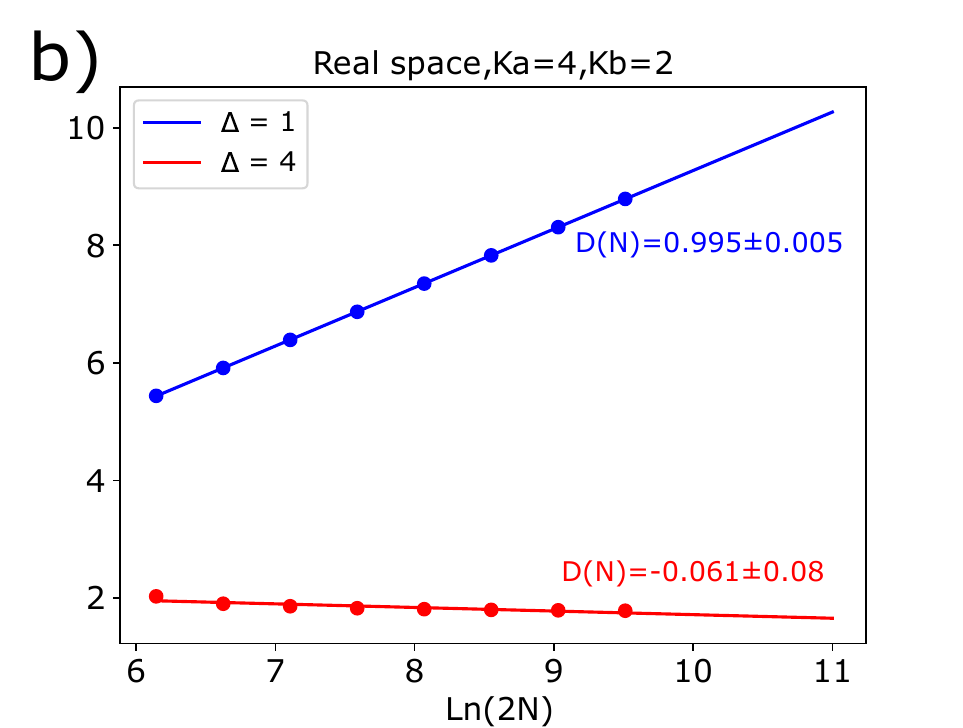}%
\end{minipage}\hspace{0.05\linewidth}%
\begin{minipage}[t]{0.45\linewidth}%
\includegraphics[scale=0.3]{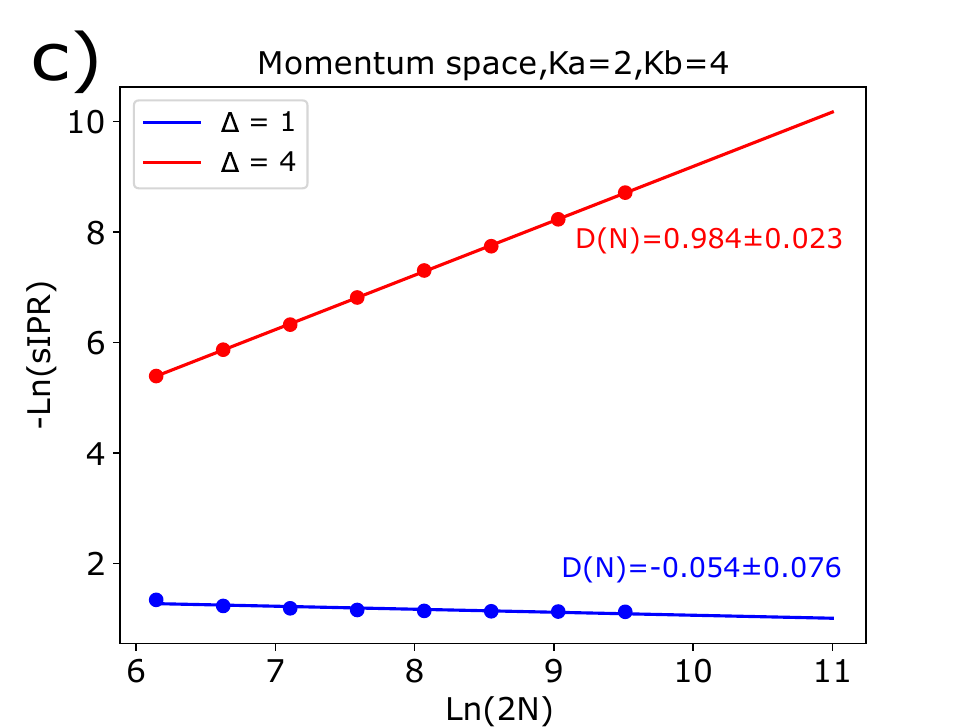}%
\end{minipage}\hspace{0.07\linewidth}%
\begin{minipage}[t]{0.45\linewidth}%
\includegraphics[scale=0.3]{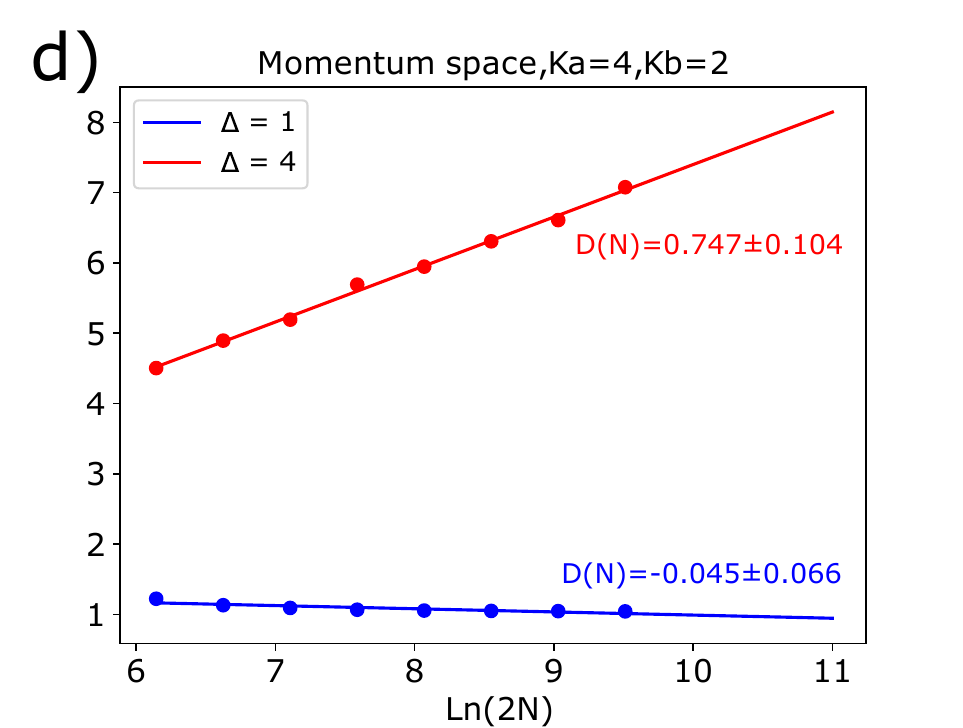}%
\end{minipage}
\caption{Plot of -Ln(sIPR) versus Ln(2N) for $\Delta = 1$ (blue curves), for $\Delta = 4$ (red curves) and for the following configurations $\mathbf{a)}$ $K_{b}>K_{a}$ in real space; $\mathbf{b)}$ $K_{a}>K_{b}$
in real space; $\mathbf{c)}$ $K_{b}>K_{a}$ in momentum space; $\mathbf{d)}$ $K_{a}>K_{b}$
in momentum space. The slope of the curves correspond to the quantity $D(N)$, a quantity that converges to the fractal dimension $\Gamma$ in the thermodynamic limit. In cases $\mathbf{a)}$, $\mathbf{b)}$ the blue curve indicates that for $\Delta=1$ the system is in the extended phase and that for $\Delta=4$ the system is in the localized phase. In case $\mathbf{c)}$, $\Delta=1$ corresponds to the localized phase and $\Delta=4$ to the extended phase. The case $\mathbf{d)}$ differs from the others in the sense that we have  the value of $D(N)$  between 0 and 1  (see detailed discussion in the text).   }
\label{fig:gamma}
\end{figure}

 Now, we can predict the fractal dimension $\Gamma$ for a given $\Delta$.  We do so by plotting -Ln(sIPR) against Ln(2N) for specific values of $\Delta$ and different system sizes. The resulting curve is linear, with slope $D(N)$, which, we stress, is the scaling exponent of the average sIPR. As $N$ increases, the slope tends to $\Gamma=1$ for the extended phase, $\Gamma=0$ for the localized phase and to $1 > \Gamma > 0$ for the critical phase. These curves are shown in Fig. \ref{fig:gamma} with the corresponding slope and confidence interval. For each case, we calculated -Ln(sIPR) for $\Delta=1$ and $\Delta=4$ and for $N = 233,377,610,987,1597,2584,4181,6765$ unit cells. In Fig. \ref{fig:gamma} (a),(b) $\Delta = 1$ corresponds to the extended phase and $\Delta = 4$ to the localized phase, just as we would qualitatively expect from Fig. \ref{fig:D(N)} (a),(b)   [and from Fig. \ref{fig:painted_IPR_diagrams} (a), (b)].   We also expect that, if more points (corresponding to greater $N$) are added to the plot, the slope of the blue curve will become progressively closer to 1 and the slope of the red curve will become 0. In Fig. \ref{fig:gamma} (c), $\Delta = 1$ corresponds to the localized phase and $\Delta = 4$ to the extended phase. In Fig. \ref{fig:gamma} (d) the value for the slope of the red curve is in agreement with the result in Fig. \ref{fig:D(N)} (d), which shows that for $\Delta = 4$, the system average scaling dimension, with $D(N) \approx 0.75$,  is dominated by the critical states.  At first glace it may seem strange that the slopes of the red curves in Fig.  \ref{fig:gamma} (b) and (d) apparently  leads to different conclusions
for $\Delta=4$, where we have $D(N)\approx 0$ in real space and $D(N)\approx 0.75$ in momentum space. 
This happens because of the same role played by the critical states
in real space and in momentum space [see Fig. \ref{fig:painted_IPR_diagrams} (b) and (d)].   In real space, and for $\Delta=4$, we have the coexistence  of localized and critical 
states [see Fig. \ref{fig:painted_IPR_diagrams} (b)] and in momentum space, also for $\Delta=4$, we have the coexistence of extended and critical states [see Fig. \ref{fig:painted_IPR_diagrams} (d)]. 
The critical states IPR scales with $N^{-\Gamma_c}$ and $N^{-\Gamma_{c^k}}$ in real and momentum space, respectively, with $0 < \Gamma_c,\Gamma_{{c^k}} < 1$ \cite{Kramer_1993,Ting2022}.  Since extended and localized states scale with exponents $\Gamma = 1$ and $\Gamma = 0$,  respectively,   we will have a
finite $D(N) = \Gamma_{{c^k}}$ value in momentum space and $D(N)\approx 0$ in real space.

\subsection{Calculation of the mobility edge}

The mobility edge (ME) is another property of quasi-periodic models that is well-known
in the literature of electronic systems \citep{Lu2022,DasSarma1,DasSarma2_ganeshan,Wang_mosaic,Wang_duality}.
This property refers to the existence of a boundary separating extended from localized states.  MEs exist in quasi-periodic one dimensional systems, 
unlike in models subjected
to Anderson disorder in one and two dimensions. Here we present
an analytical derivation of MEs,  which was previously thought not to be possible \cite{Tong2022},  based on Avila's global theory \citep{Avila}.
We note, however, that this derivation is only valid for the original,
non-chiral version of our model, as in Eq. (\ref{eq_motion_AA}) setting
$\begin{array}{c}
K_{0,j}^{A}=K_{0,j}^{B}=0\end{array}$. This is due to the fact that, in the chiral version, the imposed
local spring removes the eigenvalue dependence in the diagonal of
the dynamical matrix, making it impossible to obtain the localization
length as a function of the eigenvalues $m\omega^{2}$ in analytical
form (see, however, the methods used in Ref.  \cite{MiguelJesus2023}). To obtain the MEs analytically, we begin with the non-chiral
equations of motion

\begin{widetext}
\begin{equation}
\begin{array}{c}
\begin{array}{c}
m\omega^{2}u_{A}^{j}=(K_{a}+K_{b}+\Delta\mathrm{cos}(2\pi Qj+\phi))u_{A}^{j}-K_{a}u_{B}^{j}-(K_{b}+\Delta\mathrm{cos}(2\pi Qj+\phi))u_{B}^{j-1}\end{array}\\
\\
m\omega^{2}u_{B}^{j}=(K_{a}+K_{b}+\Delta\mathrm{cos}(2\pi Q[j+1]+\phi))u_{B}^{j}-K_{a}u_{A}^{j}-(K_{b}+\Delta\mathrm{cos}(2\pi Q[j+1]+\phi))u_{A}^{j+1}
\end{array}
\end{equation}
where $j\in[1,N]$ is the unit cell index. Then, we isolate $u_{B}^{j}$
and $u_{B}^{j-1}$as a function of $u_{A}^{j}$ and $u_{A}^{j+1}$
\begin{equation}
\begin{array}{c}
\begin{array}{c}
u_{B}^{j}=\frac{K_{a}u_{A}^{j}+(K_{b}+\Delta\mathrm{cos}(2\pi Q[j+1]+\phi))u_{A}^{j+1}}{K_{a}+K_{b}-m\omega^{2}+\Delta\mathrm{cos}(2\pi Q[j+1]+\phi)}\end{array}\\
\\
u_{B}^{j-1}=\frac{K_{a}u_{A}^{j-1}+(K_{b}+\Delta\mathrm{cos}(2\pi Qj+\phi))u_{A}^{j}}{K_{a}+K_{b}-m\omega^{2}+\Delta\mathrm{cos}(2\pi Qj+\phi)}
\end{array}
\end{equation}
For cleaner notation, we define $Z=m\omega^{2}-K_{a}-K_{b}$ and $f_{j}=\Delta\mathrm{cos}(2\pi Qj+\phi)$.
Substituting $u_{B}^{j}$ and $u_{B}^{j-1}$ in the equation for $u_{A}^{j}$,
we obtain

\begin{equation}
\begin{array}{c}
m\omega^{2}u_{A}^{j}=(K_{a}+K_{b}+f_{j})u_{A}^{j}-K_{a}\left(\frac{K_{a}u_{A}^{j}+(K_{b}+f_{j+1})u_{A}^{j+1}}{f_{j+1}-Z}\right)-(K_{b}+f_{j})\left(\frac{K_{a}u_{A}^{j-1}+(K_{b}+f_{j})u_{A}^{j}}{f_{j}-Z}\right)\end{array}
\end{equation}
Rearranging terms it is easy to see that
\begin{equation}
\begin{array}{c}
u_{A}^{j+1}=\frac{(Z-f_{j})^{2}(Z-f_{j+1})-K_{a}^{2}(Z-f_{j})-(K_{b}+f_{j})^{2}(Z-f_{j+1})}{K_{a}(K_{b}+f_{j+1})(Z-f_{j})}u_{A}^{j}-\frac{K_{a}(K_{b}+f_{j})(Z-f_{j+1})}{K_{a}(K_{b}+f_{j+1})(Z-f_{j})}u_{A}^{j-1}\end{array}\label{eq:uaj_plus_1}
\end{equation}
Now, with Eq. (\ref{eq:uaj_plus_1}) we can define a transfer matrix
$T^{j}$ as
\begin{equation}
\begin{array}{c}
\left[\begin{array}{c}
u_{A}^{j+1}\\
u_{A}^{j}
\end{array}\right]=T^{j}\left[\begin{array}{c}
u_{A}^{j}\\
u_{A}^{j-1}
\end{array}\right],\\
\begin{array}{c}
\end{array}\\
T^{j}(\omega,\phi)=\left[\begin{array}{cc}
\frac{(Z-f_{j})^{2}(Z-f_{j+1})-K_{a}^{2}(Z-f_{j})-(K_{b}+f_{j})^{2}(Z-f_{j+1})}{K_{a}(K_{b}+f_{j+1})(Z-f_{j})} & -\frac{K_{a}(K_{b}+f_{j})(Z-f_{j+1})}{K_{a}(K_{b}+f_{j+1})(Z-f_{j})}\\
1 & 0
\end{array}\right]
\end{array}
\end{equation}
The maximal Lyapunov exponent $\gamma(\omega)$ can be calculated
by computing the limit \citep{pollicott2010}
\begin{equation}
\gamma(\omega)=\lim_{N\rightarrow\infty}\frac{1}{N}{\rm Ln}||\prod_{j=1}^{N}T^{j}||_{2}
\end{equation}
where $\rm{Ln}||{\rm M}||_{2}$ denotes the natural logarithm of the spectral
radius of the matrix ${\rm M}$, i.e the maximum of the absolute values
of its eigenvalues \citep{pollicott2010}. We proceed by writing
$T^{j}$ as a product
\begin{equation}
\begin{array}{c}
T^{j}(\omega,\phi)=A^{j}(\omega,\phi)B^{j}(\omega,\phi),\\
\\
\begin{array}{c}
\begin{array}{c}
A^{j}(\omega,\phi)=\frac{K_{a}^{2}}{K_{a}(K_{b}+f_{j+1})},\end{array}\end{array}\\
\\
B^{j}(\omega,\phi)=\left[\begin{array}{cc}
\frac{(Z-f_{j})^{2}(Z-f_{j+1})-K_{a}^{2}(Z-f_{j})-(K_{b}+f_{j})^{2}(Z-f_{j+1})}{K_{a}^{2}(Z-f_{j})} & -\frac{K_{a}(K_{b}+f_{j})(Z-f_{j+1})}{K_{a}^{2}(Z-f_{j})}\\
\frac{K_{a}(K_{b}+f_{j+1})}{K_{a}^{2}} & 0
\end{array}\right]
\end{array}
\end{equation}

\end{widetext}

Note that all terms must be dimensionless, a multiplication by $K_{a}^{2}$ is required
in $A^{j}$ and a division by $K_{a}^{2}$ is required in $B^{j}$. Using these elements, we
seek to calculate the Lyapunov exponent: $\gamma(\omega)=\gamma_{A}(\omega)+\gamma_{B}(\omega)$,
where 
\begin{equation}
\begin{array}{c}
\gamma_{A}(\omega)=\lim_{N\rightarrow\infty}\frac{1}{N}{\rm Ln}\prod_{j=1}^{N}\frac{K_{a}^{2}}{\left|K_{a}(K_{b}+\Delta\mathrm{cos}(2\pi Q[j+1]+\phi))\right|},\\
\\
\begin{array}{c}
\begin{array}{c}
\gamma_{B}(\omega)=\lim_{N\rightarrow\infty}\frac{1}{N}{\rm Ln}||\prod_{j=1}^{N}B^{j}(\omega,\phi)||_{2}\end{array}\end{array}\\
\\
\end{array}
\end{equation}

For $\gamma_{A}(\omega)$ we apply ergodic theory, as usual in this
type of calculation \citep{Lu2022,Wang_duality}. This allows us
to write $\gamma_{A}(\omega)$ as an integral over the phase $\varphi$:

\small

\begin{equation}
\begin{array}{c}
\gamma_{A}(\omega)=\frac{1}{2\pi}\intop_{0}^{2\pi}{\rm Ln}\left|\frac{K_{a}^{2}}{K_{a}(K_{b}+\Delta\mathrm{cos}(\varphi))}\right|d\varphi\\
\begin{array}{c}
\end{array}\\
=\left\{ \begin{array}{c}
-{\rm Ln}\left(\frac{K_{b}+\sqrt{K_{b}^{2}-\Delta^{2}}}{2K_{a}}\right)\;,K_{b}>\Delta\\
-{\rm Ln}\left(\frac{\Delta}{2K_{a}}\right)\;,K_{b}<\Delta
\end{array}\right.
\end{array}
\\
\\
\\
\end{equation}

\normalsize

As for $\gamma_{B}(\omega)$, we start by complexifying the phase
of the cosine terms in $B^{j}(\omega,\phi)$: $\Delta\mathrm{cos}(2\pi Qj+\phi)\rightarrow\Delta\mathrm{cos}(2\pi Qj+\phi+i\epsilon)$.
As $\epsilon\rightarrow\infty$, $f_{j}(\epsilon)=\Delta\mathrm{cos}(2\pi Qj+\phi+i\epsilon)\rightarrow\frac{\Delta}{2}e^{-(2\pi iQj+i\phi)}e^{\epsilon}$,
since the exponential proportional to $e^{-\epsilon}$ vanishes. In
this scenario it is easy to see that
\begin{equation}
\frac{(Z-f_{j+1}(\epsilon))}{(Z-f_{j}(\epsilon))}\overset{\epsilon\rightarrow\infty}{\longrightarrow}\frac{\frac{\Delta}{2}e^{-(2\pi iQ[j+1]+i\phi)}e^{\epsilon}}{\frac{\Delta}{2}e^{-(2\pi iQj+i\phi)}e^{\epsilon}}=e^{-2\pi iQ}
\end{equation}

Then, the matrix $B^{j}(\omega,\phi,\epsilon)$ becomes

\begin{widetext}
\begin{equation}
\\
\\
B^{j}(\omega,\phi,\epsilon)=\left[\begin{array}{cc}
\frac{(Z-f_{j}(\epsilon))^{2}e^{-2\pi iQ}-K_{a}^{2}-(K_{b}+f_{j}(\epsilon))^{2}e^{-2\pi iQ}}{K_{a}^{2}} & -\frac{(K_{b}+f_{j}(\epsilon))e^{-2\pi iQ}}{K_{a}}\\
\frac{(K_{b}+f_{j+1}(\epsilon))}{K_{a}} & 0
\end{array}\right]
\end{equation}

\end{widetext}

\begin{figure}
\begin{raggedright}
\begin{minipage}[t]{0.45\linewidth}%
\includegraphics[scale=0.30]{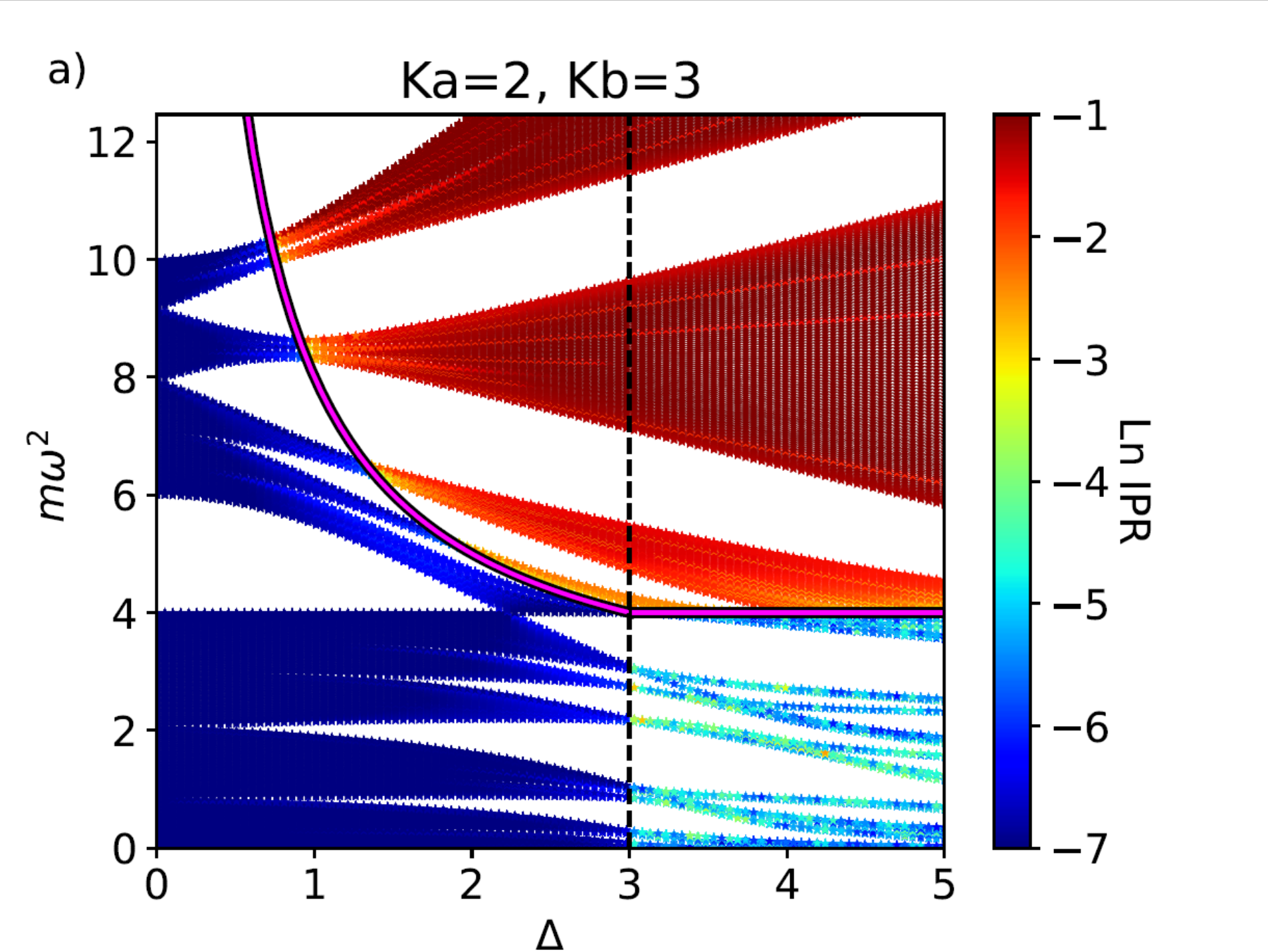}%
\end{minipage}\hspace{0.3\linewidth}%
\begin{minipage}[t]{0.45\linewidth}%
\includegraphics[scale=0.30]{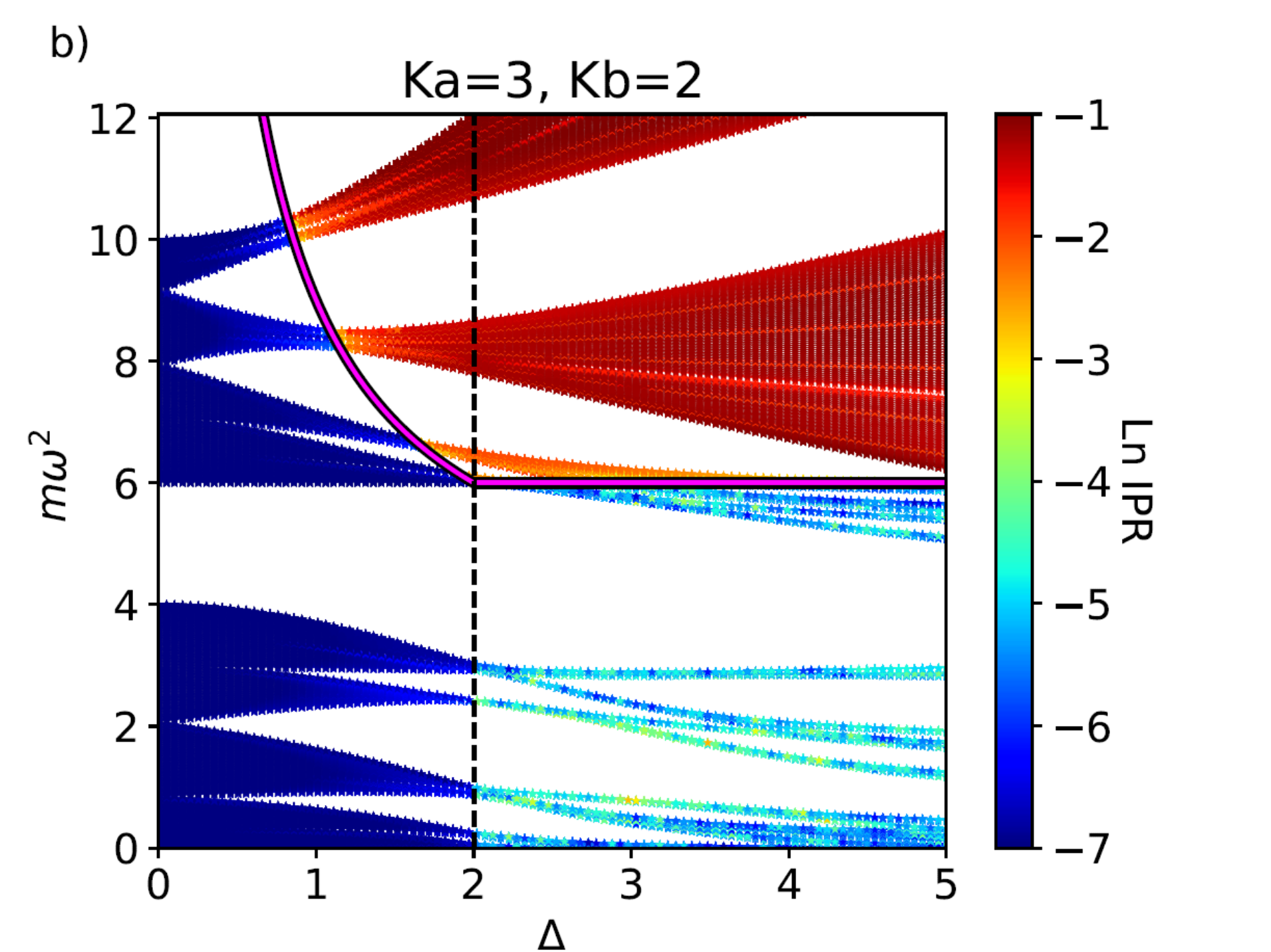}%
\end{minipage}\caption{Ln(IPR) as a function of the eigenvalues $m\omega^{2}$ for the non-chiral
version of our model and $\mathbf{a)}$$K_{b}>K_{a}$ and $\mathbf{b)}$$K_{a}>K_{b}$.
The pink line is the result of our analytical calculation and precisely
predicts the border between localized and non-localized eigenmodes.
The black dashed line corresponds to $\Delta=K_{b}$. To the left
of this dashed line, the MEs correspond to $m\omega^{2}=\frac{K_{a}(K_{b}+\Delta)}{\Delta}$
and to the right of it, to $m\omega^{2}=2K_{a}$. }
\label{fig: MEs_analitic}
\par\end{raggedright}
\end{figure}

Working with each term explicitly, substituting $Z=m\omega^{2}-K_{a}-K_{b}$,$f_{j}(\epsilon)=\frac{\Delta}{2}e^{-(2\pi iQj+i\phi)}e^{\epsilon}$
and keeping only the highest power in $\epsilon$ 
\begin{equation}
B^{j}(\omega,\phi,\epsilon)=e^{-(2\pi iQ[j+1]+i\phi)}e^{\epsilon}\left[\begin{array}{cc}
\frac{-\Delta(m\omega^{2}-K_{a})}{K_{a}^{2}} & -\frac{\Delta}{2K_{a}}\\
\frac{\Delta}{2K_{a}} & 0
\end{array}\right].
\end{equation}
The Lyapunov exponent $\gamma_{B}(\omega,\epsilon)$ is the natural
logarithm of the spectral norm of $B^{j}(\omega,\phi,\epsilon)$
\begin{widetext}
\begin{equation}
\gamma_{B}(\omega,\epsilon)={\rm Ln}||B^{j}(\omega,\phi,\epsilon)||_{2}=\epsilon+{\rm Ln}\left|\frac{\Delta\left(K_{a}-m\omega^{2}+\sqrt{m\omega^{2}(m\omega^{2}-2K_{a})}\right)}{2K_{a}^{2}}\right|
\end{equation}
\end{widetext}
 By the global theory \citep{Avila}, $\gamma_{B}(\omega)={\rm Ln}\left|\frac{\Delta\left(K_{a}-m\omega^{2}+\sqrt{m\omega^{2}(m\omega^{2}-2K_{a})}\right)}{2K_{a}^{2}}\right|$.
Then, the Lyapunov exponent is
\begin{equation}
\gamma(\omega)=\left\{ \begin{array}{c}
{\rm Ln}\left|\frac{\Delta\left(K_{a}-m\omega^{2}+\sqrt{m\omega^{2}(m\omega^{2}-2K_{a})}\right)}{K_{a}(K_{b}+\sqrt{K_{b}^{2}-\Delta^{2}})}\right|\;,K_{b}>\Delta\\
\begin{array}{c}
\end{array}\\
{\rm Ln}\left|\frac{\left(K_{a}-m\omega^{2}+\sqrt{m\omega^{2}(m\omega^{2}-2K_{a})}\right)}{K_{a}}\right|\;,K_{b}<\Delta
\end{array}\right.
\end{equation}
At the MEs, the localization length $\Lambda$ diverges \citep{Wang_duality}.
Since $\Lambda=\gamma^{-1}(\omega)$, the condition for obtaining
the MEs is $\gamma(\omega)=0.$ Thus, solving for $m\omega^{2}$ yields
\begin{equation}
m\omega^{2}=\left\{ \begin{array}{c}
\frac{K_{a}(K_{b}+\Delta)}{\Delta}\;,K_{b}>\Delta\\
\begin{array}{c}
\end{array}\\
2K_{a}\;,K_{b}<\Delta
\end{array}\right.
\end{equation}
In Fig. \ref{fig: MEs_analitic} we match our analytical results with
the diagram of Ln(IPR) as a function of $\Delta$ and $m\omega^{2}$.
The diagram was calculated for a chain with 987 unit cells. One interesting
property of this model is that all three localization regimes coexist
within very well defined boundaries. For all cases studied, we note
that above the mobility edge curve (pink curve in Fig. \ref{fig: MEs_analitic})
all eigenstates are localized; Bellow the curve, if $\Delta<K_{b}$,
all states are extended and if $\Delta>K_{b}$, all states are critical.
Furthermore, critical states are always located below the threshold
value $m\omega^{2}=2K_{a}$, such that the rectangle defined by $\Delta>K_{b},m\omega^{2}=2K_{a}$
always predicts the location of critical eigenmodes. 

\section{Final remarks}

In this work we studied topological transitions and localization properties of
a mechanical SSH model with intercell spring-constants subjected to
an Aubry-André modulation. We applied an additional local spring on each
mass to ensure chiral symmetry and used a topological invariant obtained
from the real-space eigenmodes to understand and visualize the
topological phases. An analytical computation of the Lyapunov exponents
allowed us
to predict for which values of the spring-constants $K_{a},K_{b}$
and Aubry-André amplitude $\Delta$ the 1D topological phase
transitions occur.  The analytical result confirmed the numerical calculation of the boundary between phases.
Additionally,  with a detailed  analysis
in both real- and momentum-spaces using the state IPRs, we
were able to associate each eigenmode of the chiral version of our
model with a given localization regime as a function of $\Delta$,
showing that it supports extended, localized and critical states.
Also, averaging all state IPRs and computing the fractal dimension
$D(N)$ allows us to study the localization properties of the system
as a whole and perform consistent finite scaling analysis of quasi-periodic
models. 

As for the non-chiral version of our model, we calculated the
mobility edges analytically for a SSH-based system, perfectly predicting
the border between localized and non-localized (either extended or critical) eigenstates. 
However, because of the rectangular boundaries bounding the critical phase, the system can only present
the coexistence of extended and localized states or, alternatively,  critical and localized states.
We also
observe that, for this model, it is possible to consistently predict
the localization regime of each eigenstate based solely on the value
of the tunable parameters $K_{a}$,  $K_{b}$, and $\Delta$. Our results
extend what is know in quantum electronic topological systems and
quantum electronic localization theory to the classical realm.

\begin{acknowledgments}
The authors thank Eduardo Castro and Miguel Gon{\c c}alves for helpful
discussions, and Miguel Gon{\c c}alves for a critical reading of the manuscript. 
The authors acknowledge support by the Portuguese Foundation
for Science and Technology (FCT) in the framework of the project PTDC/FIS-MAC/2045/2021.
T.V.C.A. acknowledges the computational resources provided by the Aalto Science-IT project.
N.M.R.P acknowledges support by FCT in the framework of the Strategic Funding UIDB/04650/2020,
COMPETE 2020, PORTUGAL 2020, FEDER, and through project EXPL/FISMAC/0953/
2021. N. M. R. P. also acknowledges the Independent Research Fund
Denmark (grant no. 2032- 00045B) and the Danish National Research
Foundation (Project No. DNRF165).
\end{acknowledgments}

\appendix
\section{ Periodic boundary conditions and rational approximants}

As stated in the main text, true quasi-periodicity can only be achieved
in infinite systems. For finite size scaling analysis, we impose periodic
boundary conditions to mimic an infinite system. This also allows
us to perform transformations between real and momentum spaces. PBC
requires the first and last spring-constant to be equal ($\phi$ was
set to 0 in this derivation for simplicity)
\begin{equation}
K_{b}+\Delta{\rm cos}(2\pi Q)=K_{b}+\Delta{\rm cos}(2\pi Q[N+1])\label{eq:cond_PBC_AA-1}
\end{equation}
If the the Aubry-André potential carried a truly irrational $Q$,
this condition would never be satisfied because the cosine would never
repeat itself. Then, it is clear that we need a rational approximant
of $Q$, defined in such a way that Eq. (\ref{eq:cond_PBC_AA-1})
is satisfied. The inverse of the golden ratio can be approximated
by dividing two consecutive numbers of the Fibonacci sequence: $\tilde{Q}=\frac{F_{n-1}}{F_{n}}$,
where $\tilde{Q}$ is the rational approximant such that the greater
the index $n$, the closer $\tilde{Q}$ is from $Q$. Now, we can
expand the RHS of Eq. (\ref{eq:cond_PBC_AA-1}) and try to find a
condition for $\tilde{Q}$ 
\begin{equation}
{\rm cos}(2\pi\tilde{Q})={\rm cos}(2\pi\tilde{Q}N){\rm cos}(2\pi\tilde{Q})-{\rm sin}(2\pi\tilde{Q}N){\rm sin}(2\pi\tilde{Q})
\end{equation}
Now, any choice of the type $\tilde{Q}=\frac{F_{n-1}}{F_{n}}=\frac{qs}{2N},s=1,2,3,...$
makes the term ${\rm sin}(2\pi\tilde{Q}N)$ vanish. This means that
we have to choose a rational approximant whose denominator $F_{n}$
is related to the size of the system. The constant $s$ is still to
be determined though. Proceeding, we have
\begin{equation}
1={\rm cos}(2\pi\tilde{Q}N)={\rm cos}(2\pi\left[\frac{qs}{2N}\right]N)\label{eq:cond_Q_AA-1}
\end{equation}
Naturally, quasi-periodicity requires no repetition of the cosine
arguments throughout the chain. Because of this we need the smallest
value of $Q$ that satisfies Eq. (\ref{eq:cond_Q_AA-1}). If $s=1$,
some approximants with odd numerator ($F_{n-1})$ would result in
$1={\rm cos}(\pi q)$, which is false. Then, we need $s=2$ such that,
independent of the Fibonacci numbers chosen, we get a cosine equal to 1
\begin{equation}
{\rm cos}(2\pi\tilde{Q}N)={\rm cos}(2\pi\left[\frac{qs}{2N}\right]N)\lfloor_{s=2}={\rm cos}(2\pi q)=1
\end{equation}
This means that the rational approximant of choice has to be $\tilde{Q}=\frac{F_{n-1}}{F_{n}}=\frac{q}{N},$where
$N$ is the number of unit cells in the system and $q$ its predecessor
in the Fibonacci sequence. Thus, the size of the system is determined
by the rational approximant chosen and vice-versa. Finally, the dynamical
matrix for the PBC system with off-diagonal A-A modulation is
\begin{widetext}
\begin{equation}
\mathbf{\mathbb{M}}=\frac{1}{m}\left[\begin{array}{ccccc}
K_{a}+K_{b} & -K_{a} & 0 & ... & -(K_{b}+\Delta\mathrm{cos}[2\pi Q+\phi])\\
-K_{a} & K_{a}+K_{b} & -(K_{b}+\Delta\mathrm{cos}[4\pi Q+\phi]) & ... & 0\\
0 & -(K_{b}+\Delta\mathrm{cos}[4\pi Q+\phi]) & K_{a}+K_{b} & ... & 0\\
... & ... & ... & ... & -K_{a}\\
-(K_{b}+\Delta\mathrm{cos}[2\pi Q+\phi]) & 0 & 0 & ... & K_{a}+K_{b}
\end{array}\right]_{2N\times2N}
\end{equation}

\end{widetext}

\section{Transformation to momentum space for hopping-modulated
SSH-based quasi-periodic systems}

Our goal in this section is to transform the equations

\begin{widetext}
\begin{equation}
\begin{array}{c}
m\omega^{2}u_{A}^{j}=(K_{a}+K_{b})u_{A}^{j}-K_{a}u_{B}^{j}-(K_{b}+\Delta{\rm cos}(2\pi Qj))u_{B}^{j-1}\\
m\omega^{2}u_{B}^{j}=(K_{a}+K_{b})u_{B}^{j}-K_{a}u_{A}^{j}-(K_{b}+\Delta{\rm cos}(2\pi Q[j+1]))u_{A}^{j+1}
\end{array}
\end{equation}
to momentum space.We begin by introducing the transformation
\begin{equation}
u_{\alpha}^{j}=\sum_{k}e^{2\pi iQkj}\psi_{\alpha}^{k}\label{eq:transf_mom}
\end{equation}
where $j\in[1,N]$ is the unit cell index in real space and $k\in[1,N]$
is the unit cell index in momentum space. $\alpha$ can be $A,B$,
depending on the sublattice type. 

\subsection*{Type-A mass equation}

We begin by transforming the equation for the A-sublattice, inserting
Eq. (\ref{eq:transf_mom}) for $\alpha=A$

\begin{equation}
m\omega^{2}u_{A}^{j}=(K_{a}+K_{b})u_{A}^{j}-K_{a}u_{B}^{j}-(K_{b}+\Delta{\rm cos}(2\pi Qj))u_{B}^{j-1}
\end{equation}
\begin{equation}
m\omega^{2}\sum_{k}e^{2\pi iQkj}\psi_{A}^{k}=(K_{a}+K_{b})\sum_{k}e^{2\pi iQkj}\psi_{A}^{k}-K_{a}\sum_{k}e^{2\pi iQkj}\psi_{B}^{k}-(K_{b}+\Delta{\rm cos}(2\pi Qj))\sum_{k}e^{2\pi iQk(j-1)}\psi_{B}^{k}
\end{equation}
\begin{equation}
=(K_{a}+K_{b})\sum_{k}e^{2\pi iQkj}\psi_{A}^{k}-\sum_{k}(K_{a}+K_{b}e^{-2\pi iQk})e^{2\pi iQkj}\psi_{B}^{k}-\Delta{\rm cos}(2\pi Qj)\sum_{k}e^{2\pi iQk(j-1)}\psi_{B}^{k}
\end{equation}
Writing the cosine as a sum of exponentials
\begin{equation}
=(K_{a}+K_{b})\sum_{k}e^{2\pi iQkj}\psi_{A}^{k}-\sum_{k}(K_{a}+K_{b}e^{-2\pi iQk})e^{2\pi iQkj}\psi_{B}^{k}-\frac{\Delta}{2}\sum_{k}e^{2\pi iQj}e^{2\pi iQk(j-1)}\psi_{B}^{k}-\frac{\Delta}{2}\sum_{k}e^{-2\pi iQj}e^{2\pi iQk(j-1)}\psi_{B}^{k}
\end{equation}
Now, we put $e^{2\pi iQj}$ in evidence in the last two terms
\begin{equation}
=(K_{a}+K_{b})\sum_{k}e^{2\pi iQkj}\psi_{A}^{k}-\sum_{k}(K_{a}+K_{b}e^{-2\pi iQk})e^{2\pi iQkj}\psi_{B}^{k}-\frac{\Delta}{2}\sum_{k}e^{2\pi iQj(k+1)}e^{-2\pi iQk}\psi_{B}^{k}-\frac{\Delta}{2}\sum_{k}e^{2\pi iQj(k-1)}e^{-2\pi iQk}\psi_{B}^{k}
\end{equation}
Since the system is closed, we can change the index of the summation.
We then make the following transformations
\begin{equation}
\begin{array}{cc}
k+1\rightarrow k', & \sum_{k=1}^{N}\rightarrow\sum_{k'=0}^{N-1}\\
k-1\rightarrow k'', & \sum_{k=1}^{N}\rightarrow\sum_{k''=2}^{N+1}
\end{array}
\end{equation}
But due to PBC, the $0^{th}$ cell is the last one and the $N+1^{th}$
cell is the first one. Then, the summations over $m'$ and $m''$
are equivalent to summations in $m$. Applying this reasoning we obtain
\begin{equation}
=(K_{a}+K_{b})\sum_{k}e^{2\pi iQkj}\psi_{A}^{k}-\sum_{k}(K_{a}+K_{b}e^{-2\pi iQk})e^{2\pi iQkj}\psi_{B}^{k}-\frac{\Delta}{2}\sum_{k}e^{2\pi iQjk}e^{-2\pi iQ(k-1)}\psi_{B}^{k-1}-\frac{\Delta}{2}\sum_{k}e^{2\pi iQjk}e^{-2\pi iQ(k+1)}\psi_{B}^{k+1}
\end{equation}
Lastly, the terms of the sum that multiply $e^{2\pi iQkj}$ have to
be equal. We then obtain the final result

\begin{equation}
m\omega^{2}\psi_{A}^{k}=(K_{a}+K_{b})\psi_{A}^{k}-(K_{a}+K_{b}e^{-2\pi iQk})\psi_{B}^{k}-\frac{\Delta}{2}e^{-2\pi iQ(k-1)}\psi_{B}^{k-1}-\frac{\Delta}{2}e^{-2\pi iQ(k+1)}\psi_{B}^{k+1}
\end{equation}
This is the result for the type-A mass equation. 

\subsection*{Type-B mass equation}

For the B-sublattice, inserting Eq. (\ref{eq:transf_mom}) for $\alpha=B$

\begin{equation}
m\omega^{2}u_{B}^{j}=(K_{a}+K_{b})u_{B}^{j}-K_{a}u_{A}^{j}-(K_{b}+\Delta{\rm cos}(2\pi Q[j+1]))u_{A}^{j+1}
\end{equation}

\begin{equation}
m\omega^{2}\sum_{k}e^{2\pi iQkj}\psi_{B}^{k}=(K_{a}+K_{b})\sum_{k}e^{2\pi iQkj}\psi_{B}^{k}-K_{a}\sum_{k}e^{2\pi iQkj}\psi_{A}^{k}-(K_{b}+\Delta{\rm cos}(2\pi Q[j+1]))\sum_{k}e^{2\pi iQk(j+1)}\psi_{A}^{k}
\end{equation}
\begin{equation}
=(K_{a}+K_{b})\sum_{k}e^{2\pi iQkj}\psi_{B}^{k}-\sum_{k}(K_{a}+K_{b}e^{2\pi iQ})e^{2\pi iQkj}\psi_{A}^{k}-\Delta{\rm cos}(2\pi Q[j+1])\sum_{k}e^{2\pi iQk(j+1)}\psi_{A}^{k}
\end{equation}
Expanding the cosine as the sum of two exponentials, we get
\begin{equation}
=(K_{a}+K_{b})\sum_{k}e^{2\pi iQkj}\psi_{B}^{k}-\sum_{k}(K_{a}+K_{b}e^{2\pi iQ})e^{2\pi iQkj}\psi_{A}^{k}-\frac{\Delta}{2}\sum_{k}e^{2\pi iQ(j+1)}e^{2\pi iQk(j+1)}\psi_{A}^{k}-\frac{\Delta}{2}\sum_{k}e^{-2\pi iQ(j+1)}e^{2\pi iQk(j+1)}\psi_{A}^{k}
\end{equation}
Now, we group exponentials with $j$ in the exponent
\begin{equation}
=(K_{a}+K_{b})\sum_{k}e^{2\pi iQkj}\psi_{B}^{k}-\sum_{k}(K_{a}+K_{b}e^{2\pi iQ})e^{2\pi iQkj}\psi_{A}^{k}-\frac{\Delta}{2}\sum_{k}e^{2\pi iQj(k+1)}e^{2\pi iQ(k+1)}\psi_{A}^{k}-\frac{\Delta}{2}\sum_{k}e^{2\pi iQj(k-1)}e^{2\pi iQ(k-1)}\psi_{A}^{k}
\end{equation}
The change in the summation index is performed exactly as for the
A-type equation
\begin{equation}
=(K_{a}+K_{b})\sum_{k}e^{2\pi iQkj}\psi_{B}^{k}-\sum_{k}(K_{a}+K_{b}e^{2\pi iQ})e^{2\pi iQkj}\psi_{A}^{k}-\frac{\Delta}{2}\sum_{k}e^{2\pi iQjk}e^{2\pi iQk}\psi_{A}^{k-1}-\frac{\Delta}{2}\sum_{k}e^{2\pi iQjk}e^{2\pi iQk}\psi_{A}^{k+1}
\end{equation}
Lastly, the terms of the sum that multiply $e^{2\pi iQkj}$ have to
be equal
\begin{equation}
m\omega^{2}\psi_{B}^{k}=(K_{a}+K_{b})\psi_{B}^{k}-(K_{a}+K_{b}e^{2\pi iQ})\psi_{A}^{k}-\frac{\Delta}{2}e^{2\pi iQk}\psi_{A}^{k-1}-\frac{\Delta}{2}e^{2\pi iQk}\psi_{A}^{k+1}.
\end{equation}
We then obtain the final result

\begin{equation}
\begin{array}{c}
m\omega^{2}\psi_{A}^{k}=(K_{a}+K_{b})\psi_{A}^{k}-(K_{a}+K_{b}e^{-2\pi iQk})\psi_{B}^{k}-\frac{\Delta}{2}e^{-2\pi iQ(k-1)}\psi_{B}^{k-1}-\frac{\Delta}{2}e^{-2\pi iQ(k+1)}\psi_{B}^{k+1}\\
\\
\begin{array}{c}
m\omega^{2}\psi_{B}^{k}=(K_{a}+K_{b})\psi_{B}^{k}-(K_{a}+K_{b}e^{2\pi iQ})\psi_{A}^{k}-\frac{\Delta}{2}e^{2\pi iQk}\psi_{A}^{k-1}-\frac{\Delta}{2}e^{2\pi iQk}\psi_{A}^{k+1}\end{array}
\end{array}
\end{equation}
With this set of equations, we can produce the diagrams in Fig. \ref{fig:painted_IPR_diagrams}. 

\end{widetext}


%

\end{document}